\documentclass[acmsmall,nonacm]{acmart}

\setcopyright{none}
\renewcommand\footnotetextcopyrightpermission[1]{}

\usepackage{pifont}
\usepackage{algorithm}
\usepackage{subcaption} % 
\usepackage{algorithmic}
\usepackage{tabularx}
\usepackage{graphicx}
\usepackage[table]{xcolor}
\usepackage{enumitem}
\usepackage{multirow}
\usepackage{bm} 
\usepackage{pgfplots}
\usepackage{booktabs}
\usepackage{makecell}

\usepgfplotslibrary{groupplots}
\usetikzlibrary{calc}

\usepackage[dvipsnames]{xcolor}
\usepackage{booktabs}
\usepackage{xcolor}
\definecolor{bad}{RGB}{255,0,0}
\usepackage{tikz}
\usepackage{caption}
\pgfplotsset{compat=1.17}

\usepackage{placeins}

\AtBeginDocument{%
  }

\begin{document}

\title{Generative Universal Multimodal Retrieval with Dual-role Identifiers}

\author{Kaipeng Li}
\orcid{0009-0004-4406-2956}
\affiliation{%
  \institution{Independent Researcher}
  \city{Tsukuba}
  \state{Ibaraki}
  \country{Japan}
}
\email{keeey.kk@gmail.com}

\author{Haitao Yu}
\authornote{Corresponding author.}
\orcid{0000-0002-1569-8507}
\affiliation{%
  \institution{Institute of Library,
  Information and Media Science, 
  University of Tsukuba}
  \city{Tsukuba}
  \state{Ibaraki}
  \country{Japan}
  }
\email{yuhaitao@slis.tsukuba.ac.jp}

\author{Xuanchen Zhou}
\orcid{0009-0002-1659-2592}
\affiliation{%
  \institution{College of Knowledge and Library Sciences, University of Tsukuba}
  \city{Tsukuba}
  \country{Japan}
}
\email{s2313451@u.tsukuba.ac.jp}

\renewcommand{\shortauthors}{Li et al.}

\begin{abstract}
Generative information retrieval (GIR) has emerged as a compelling alternative to the conventional \textit{index-retrieve-then-rank} retrieval pipeline by training a generator to produce the identifiers of relevant items directly. Despite its promise, a number of open challenges still remain. First, constrained left-to-right decoding is vulnerable to prefix-level errors and local optima. Second, most prior GIR research remains largely unimodal, leaving instruction-aware retrieval across text, image, and mixed image-text items underexplored. Third, although discrete identifier–based GIR offers higher efficiency, its retrieval accuracy still lags behind that of the cutting-edge dense-vector-based retrieval methods. Motivated by these challenges, we propose \textbf{DrIG}, a novel \textbf{G}enerative framework for universal multimodal retrieval featuring \textbf{D}ual-\textbf{r}ole \textbf{I}dentifiers, which supports diverse retrieval tasks across multiple modalities and domains. Each candidate is assigned a single residual-quantized identifier that serves two complementary roles. In its \emph{sequential role}, the identifier is decoded autoregressively, where the first token explicitly models modality and the remaining tokens capture progressively finer semantics. In its \emph{set-based role}, the same tokens are reinterpreted as an unordered set to provide a prefix-independent relevance prior, which guides constrained beam search and alleviates local-optimum errors. To further compensate for the information loss introduced by discretization, we incorporate a hybrid retrieval strategy that reranks the top generative candidates based on dense-vector-based similarity. Extensive experiments\footnote{The code will be released.} on the M-BEIR benchmark and the text-to-image evaluation datasets (Flickr30K and MSCOCO) show that: (1) DrIG consistently outperforms state-of-the-art generative multimodal retrieval baselines across diverse retrieval tasks. The hybrid reranking successfully combines the strengths of two paradigms and yields a favorable efficiency-effectiveness trade-off relative to strong dense baselines. (2) Comprehensive ablation and scaling analyses show that the base LMM, beam size, reranking depth, and fusion strategy significantly affect retrieval performance, providing practical guidance for future generative multimodal retrieval systems.
\end{abstract}

%    <concept>
%        <concept_id>10002951</concept_id>
%        <concept_desc>Information systems</concept_desc>
%        <concept_significance>500</concept_significance>
%        </concept>
%    <concept>
%        <concept_id>10002951.10003317.10003371.10003386</concept_id>
%        <concept_desc>Information systems~Multimedia and multimodal retrieval</concept_desc>
%        <concept_significance>500</concept_significance>
%        </concept>
%  </ccs2012>

% \ccsdesc[500]{Information systems}
% \ccsdesc[500]{Information systems~Multimedia and multimodal retrieval}

% \keywords{Universal Multimodal Retrieval, Dual-role Identifier, LMM}

\maketitle

\section{Introduction}
\label{sec:introduction}
Nowadays, information retrieval (IR) systems play a crucial role in bridging the ever-expanding World Wide Web (WWW) with diverse user information needs, supporting activities that range from fact finding and media search to decision making and knowledge access. Classical IR methods (including lexical matching, learning-to-rank, and modern dense retrieval) have substantially advanced retrieval effectiveness over the past decades~\cite{jones1972statistical,robertson1994okapi,burges2006lambdarank,khattab2020colbert,metzler2021rethinking}. Nevertheless, most of the prior methods boil down to \textit{the multi-stage index-retrieve-then-rank paradigm} \cite{metzler2021rethinking}, suffering from objective inconsistency across stages. Motivated by the recent success of large language models (LLMs), \emph{generative information retrieval} (GIR) has emerged as a promising alternative~\cite{tay2022transformer,bevilacqua2022autoregressive,li2025survey}. Instead of exhaustively scoring all candidates, GIR first maps each candidate to a discrete identifier and then trains a generator to produce the identifiers of relevant items directly. This formulation shifts retrieval from similarity search in a large candidate space to conditional generation in a compact discrete space. Towards effective GIR, many methods have been proposed. Based on the modalities involved, we can categorize relevant studies into three groups: (1) \textit{generative unimodal retrieval} (GUR), which focuses on retrieval within a single modality, such as \cite{tay2022transformer, mehta2022dsi++, wang2023novo, qiao2023diffusionret, li2024learning, zeng2024scalable, li2023multiview, sun2023learning, bevilacqua2022autoregressive, chenetal2023understanding, chen2023continual, si2024generative, pradeep-etal-2023-generative, zhuang2022bridging, wang2022neural, tang2025generative, zhang2025replication, mekonnen2025lightweight, wu2025constrained, tang2024generative, kim-etal-2024-exploring-practicality}; (2) \textit{generative cross-modal retrieval} (GCMR), which addresses retrieval across two distinct modalities, such as \cite{li2025semcore, li2024revolutionizing, hendriksen2025benchmark, qu2025tiger}; and (3) \textit{generative universal multimodal retrieval} (GUMR), which enables retrieval across multiple modalities within a unified framework, such as \cite{wei2024uniir, kim2025genius, lin2024mm}. 

A closer look at the previous studies on GIR shows that most prior studies focus on GUR, while GCMR and GUMR remain relatively underexplored. In particular, multimodal retrieval has rapidly evolved from conventional image-text matching to more general \emph{universal multimodal retrieval}, where a query may contain text, images, or both, and the target can likewise belong to different modalities and domains~\cite{cao2022survey,wei2024uniir}. This setting is particularly appealing because it provides a unified formulation for a wide range of tasks, including text-to-image retrieval, image-to-text retrieval, multimodal question answering retrieval, and multimodal evidence retrieval. Recent retrieval methods based on large multimodal models (LMMs) have demonstrated strong representation capacity in such settings~\cite{liu2025lamra,lin2024mm}. Yet these methods still inherit the core limitations of dense retrieval: they require a candidate pool to be represented and searched in continuous space, and their computational cost scales with the retrieval corpus. Applying GIR to universal multimodal retrieval is therefore attractive, but it introduces a set of nontrivial challenges. First, most generative retrievers decode identifiers from left to right with constrained beam search. This makes the retrieval result highly sensitive to early prefix decisions: once the prefix of a relevant candidate is pruned, the candidate becomes irrecoverable, even if it is globally relevant. Second, multimodal retrieval requires the model to represent modality distinctions and fine-grained semantics simultaneously. A naive identifier design may fail to encode both aspects effectively. Third, discrete identifiers inevitably compress continuous embeddings, which improves efficiency but also discards information that is often crucial for fine-grained ranking. As a result, generative multimodal retrievers typically still lag behind strong dense baselines in effectiveness.

In this work, we address these challenges with a simple but powerful idea: the same discrete identifier can serve \emph{two complementary roles}. On the one hand, it acts as an ordered token sequence that can be generated autoregressively. On the other hand, the same tokens can be reinterpreted as an unordered set that supports order-invariant relevance estimation. Building on this idea, we propose DrIG, a generative framework for universal multimodal retrieval with dual-role identifiers. DrIG first learns instruction-aware multimodal embeddings using an LMM and contrastive fine-tuning. It then converts candidate embeddings into residual-quantized identifiers whose first token explicitly captures modality while later tokens encode progressively finer semantics. During inference, sequential decoding scores and order-invariant global relevance priors are combined to guide constrained beam search, reducing the risk that relevant candidates are discarded due to locally suboptimal prefixes. To further narrow the effectiveness gap between generative and dense retrieval, DrIG also incorporates a \emph{hybrid retrieval strategy}. The generative retriever first produces a compact top-$k$ candidate list efficiently; the top-$k$ results are then reranked by continuous similarities in the original embedding space. This design preserves the scalability advantage of generative retrieval while recovering fine-grained distinctions that may be lost during quantization. In addition, we introduce query augmentation through query--target interpolation and a discriminative ranking objective for decoder training, which together improve robustness and ranking consistency. To summarize, the main contributions of this paper can be listed as follows:
\begin{itemize}
    \item We propose \textbf{DrIG}, a novel generative framework for universal multimodal retrieval that assigns each candidate a single residual-quantized identifier and reuses it in two complementary roles: a sequential role for autoregressive generation and a set-based role for prefix-independent relevance estimation.    
    \item We introduce \textit{dual-guided constrained decoding} that combines prefix-valid constrained beam search with an order-invariant global relevance prior, thereby mitigating the local-optimum problem of left-to-right identifier generation. We further  integrate DrIG with dense top-k reranking and systematically analyze the resulting effectiveness–efficiency trade-off.
    \item We provide ablations and diagnostic analyses on codebook design, beam size, prior weight, reranking depth, decoder backbone, and key training objectives, providing practical insights into the design of generative multimodal retrieval systems.
\end{itemize}

The remainder of this paper is organized as follows. Section~\ref{sec:related_work} reviews prior work on multimodal retrieval and recent advances in GIR. Section~\ref{sec:preliminaries} introduces the problem formulation of GUMR. Section~\ref{sec:method} presents the proposed DrIG framework, including instruction-aware multimodal representation learning, dual-role identifier construction, dual-guided constrained decoding, decoder training objectives, and hybrid reranking. Section~\ref{sec:experiments} describes the experimental setup and reports the main results on M-BEIR, together with additional text-to-image retrieval experiments on Flickr30K and MSCOCO. We further provide component ablations, visualization analyses, effectiveness--efficiency comparisons, hyperparameter studies, and qualitative case studies. Finally, Section~\ref{sec:conclusion} concludes the paper and discusses future research directions.

\section{Related Work}
\label{sec:related_work}
In this section, we first review non-generative multimodal retrieval methods, including shared-embedding approaches, cross-modal interaction models, and recent LMM-based methods. Due to space constraints, we refer the reader to the work~\cite{ijcai2022p759} for a detailed overview. Then we describe recent advances in GIR and position our work with respect to closely related methods.

\subsection{Non-generative Multimodal Retrieval}
Early multimodal retrieval studies mainly focused on image–text matching. A large body of work learns shared embedding spaces where images and texts can be compared directly, while another line of research introduces cross-modal interaction mechanisms to improve matching quality. Representative approaches include convolutional architectures, metric-learning objectives, cross-attention models, and Transformer-based matching networks~\cite{gong2014improving,lee2018stacked,nam2017dual,li2021align}. These methods are often instantiated in one of three architectural families. \emph{Two-tower} models encode different modalities separately and compare them with a lightweight similarity function, making them scalable for large candidate pools~\cite{radford2021learning,chen2021learning,zheng2020dual}. \emph{Fusion-based} or \emph{two-leg} models introduce cross-modal interaction to improve matching quality, although usually at a higher computational cost~\cite{singh2022flava,yu2022coca}. \emph{One-tower} models instead attempt to unify multimodal encoding within a single backbone~\cite{jang2023unifying,tschannen2023clippo}.

More recently, multimodal retrieval has been extended from fixed image--text matching to universal multimodal retrieval, where queries and candidates may be text, images, or image--text pairs, and the retrieval intent is specified by natural-language instructions. UniIR/M-BEIR~\cite{wei2024uniir} provides a benchmark covering diverse retrieval tasks, modalities, and domains. Meanwhile, LMMs have been adapted for retrieval by leveraging their stronger language understanding and multimodal reasoning capacity. Examples include instruction-aware retrieval assistants and universal multimodal embedding models~\cite{wei2024uniir,jiang2025vlm2vec,lin2024mm,liu2025lamra}. VLM2Vec~\cite{jiang2025vlm2vec} converts vision--language models into general-purpose multimodal embedders through contrastive training on massive multimodal embedding tasks. MM-Embed~\cite{lin2024mm} fine-tunes multimodal LLMs as universal multimodal retrievers and further shows that MLLM-based rerankers can improve retrieval results for complex multimodal queries. LamRA~\cite{liu2025lamra} adapts LMMs through language-only pretraining and multimodal instruction tuning, enabling both retrieval and reranking across heterogeneous multimodal tasks.

These non-generative embedding-based retrieval methods provide powerful representations. However, they still fundamentally rely on dense candidate scoring, approximate nearest-neighbor search, or expensive ranking over candidate lists. Our work complements this line of research by using LMMs to obtain strong multimodal representations, but it departs from continuous search at inference time through generative identifier decoding.

\subsection{Generative Information Retrieval}
In this section, we introduce representative studies on GIR and then clarify how DrIG differs from closely related methods. We refer the reader to the work~\cite{li2025survey} for a detailed survey on GIR.

\subsubsection{Generative Unimodal Retrieval}
Generative unimodal retrieval has been most extensively studied in text and document retrieval, where the goal is to retrieve textual documents by generating their identifiers. Representative early methods, such as DSI~\cite{tay2022transformer}, NCI~\cite{wang2022neural}, and autoregressive search engines~\cite{bevilacqua2022autoregressive}, demonstrate the feasibility of using sequence models as differentiable or generative indexes. Subsequent studies~\cite{tay2022transformer,bevilacqua2022autoregressive,wang2022neural,mehta2022dsi++,chenetal2023understanding,pradeep-etal-2023-generative,zhuang2022bridging,li2023multiview,sun2023learning,wang2023novo,si2024generative,qiao2023diffusionret,li2024learning,zeng2024scalable,zeng2024planning,tang2024generative,chen2023continual,kim-etal-2024-exploring-practicality,zhang2025replication,tang2025generative,mekonnen2025lightweight,wu2025constrained,zhang2025multilevel} have improved this paradigm from several perspectives, including identifier design, training objectives, decoding strategies, scalability, and adaptation to dynamic corpora.

A central challenge in generative retrieval is how to construct identifiers that are both easy for a sequence model to generate and sufficiently discriminative for retrieval. Existing work has explored semantic strings, learned document tokens, interpretable identifiers, multi-view identifiers, and tree-structured identifiers [34, 56, 59, 66]. Recent studies on document identifier learning further emphasize that effective identifiers should be descriptive, discriminative, and aligned with relevance signals~\cite{li2023multiview,sun2023learning,wang2023novo,si2024generative}. Another line of work focuses on improving the training and decoding objectives of generative retrievers. LTRGR~\cite{li2024learning} introduces a learning-to-rank objective to reduce the mismatch between token-level generation and document-level ranking. DiffusionRet~\cite{qiao2023diffusionret} combines diffusion modeling with constrained decoding. RIPOR~\cite{zeng2024scalable} proposes prefix-oriented ranking optimization and relevance-based identifier initialization, showing that prefix-level supervision is important for scalable generative retrieval. More recent studies investigate direct document relevance optimization~\cite{mekonnen2025lightweight}, the limitations of constrained autoregressive decoding~\cite{wu2025constrained}, and the behavior of generative retrievers under dynamic corpora~\cite{chen2023continual,kim-etal-2024-exploring-practicality,zhang2025replication}.

The aforementioned studies reveal a common limitation of purely left-to-right identifier generation: a relevant item can be irreversibly discarded once its prefix is pruned during constrained decoding. To cope with this challenge, TSGen~\cite{zhang2024termset} replaces purely sequential document identifiers with term sets and uses permutation-invariant decoding to reduce false-pruning errors caused by left-to-right generation. PAG~\cite{zeng2024planning} creates the set-based DocIDs under the bag-of-words assumption and sequential DocIDs based on the relevance-based document representations to support simultaneous and autoregressive decodings, respectively. Inspired these two studies, DrIG designs identifiers with two complementary roles: an ordered sequence to facilitate Trie-constrained autoregressive generation, and a prefix-independent set-based representation for global relevance guidance. Furthermore, DrIG targets a more heterogeneous setting. Rather than designing identifiers only for unimodal text retrieval, DrIG constructs modality-aware residual-quantized identifiers for text, image, and image–text candidates. 

\subsubsection{Generative Cross-modal Retrieval}
GCMR extends GIR beyond text-only retrieval, most commonly to text-to-image retrieval. Compared with unimodal document retrieval, this setting introduces additional challenges because the model must bridge modality gaps while still generating valid candidate identifiers. Recent methods have explored different ways to represent visual candidates as discrete identifiers and train multimodal generators to retrieve them~\cite{li2024generative,zhang2024irgen,qu2025tiger,li2024revolutionizing,fang2025cart,licomgtir,li2025semcore}. GRACE~\cite{li2024generative} assigns identifier strings to images and trains a multimodal language model to memorize and retrieve images by generating their identifiers. IRGen~\cite{zhang2024irgen} recasts image retrieval as a sequence-to-sequence generative modeling problem. TIGeR~\cite{qu2025tiger} explores the connection between text-to-image generation and retrieval, using large multimodal models to unify the two tasks. AVG~\cite{li2024revolutionizing} formulates text-to-image retrieval as autoregressive token-to-visual-token generation. CART~\cite{fang2025cart} proposes a coarse-to-fine generative cross-modal retrieval framework that combines clustering and residual vector quantization to construct multimodal identifiers. ComGTIR~\cite{licomgtir} further introduces dual identifiers and hybrid retrieval strategies for generative text-image retrieval, using a sequential identifier and an order-invariant identifier to mitigate local-optimum errors during constrained beam search.

These methods demonstrate that generative retrieval can be successfully extended from text-only retrieval to cross-modal scenarios. However, most existing GCMR methods are designed for task-specific retrieval, especially text-to-image retrieval, where the query modality and target modality are predefined. DrIG differs from this line of work in both problem setting and identifier design. First, DrIG targets universal multimodal retrieval, where queries and candidates may be text, image, or image–text pairs, and the target modality must be inferred from the instruction. Second, DrIG does not maintain separate identifier spaces for sequential generation and global matching. Instead, it assigns each candidate a single residual-quantized identifier and reuses the same code tokens as both an ordered sequence and a prefix-independent set-based relevance representation.

\subsubsection{Generative Universal Multimodal Retrieval}
GUMR is an emerging direction that aims to support retrieval across heterogeneous query and candidate modalities within a unified generative framework. This setting is more challenging than conventional text-to-image retrieval because the model must jointly infer semantic relevance, target modality, and task intent from the input instruction. Wei et al.~\cite{wei2024uniir} establish an important benchmark for universal multimodal retrieval by assembling diverse datasets across multiple tasks, modalities, and domains. However, they mainly focus on dense vector-based retrieval rather than generative identifier decoding.

The closest prior work to DrIG is GENIUS~\cite{kim2025genius}, which proposes a generative framework for universal multimodal search. GENIUS introduces modality-decoupled semantic quantization to transform multimodal candidates into discrete identifiers and uses query augmentation to improve generalization across diverse query forms. Despite its effectiveness, GENIUS still primarily relies on sequential identifier generation. As a result, it remains vulnerable to prefix-level pruning errors during constrained beam search. Once the prefix of a relevant candidate is discarded, the candidate cannot be recovered in subsequent decoding steps. DrIG addresses this limitation by introducing dual-role identifiers for generative universal multimodal retrieval. This design preserves the validity and efficiency of constrained generative retrieval while reducing the risk of local-optimum errors caused by early prefix pruning.

\section{Preliminaries}
\label{sec:preliminaries}
In \emph{universal multimodal retrieval} \cite{wei2024uniir}, a query is formulated as a pair
$\widehat{q}=(q_{\mathrm{con}}, q_{\mathrm{inst}})$, where $q_{\mathrm{con}}$ denotes the query content and
$q_{\mathrm{inst}}$ is a natural-language instruction. The query content may be text $q_t$, image $q_i$, or a multimodal pair $(q_i,q_t)$. The instruction specifies the retrieval objective, including the desired target modality and the application domain. A candidate item $\widehat{c}$ can likewise be text $c_t$, image $c_i$, or an image--text pair $(c_i,c_t)$. Under this formulation, a single retrieval model can support diverse tasks such as text-to-image, image-to-text, multimodal-to-text, and multimodal-to-multimodal retrieval in a unified manner.

In \emph{generative universal multimodal retrieval} (GUMR), a retrieval system typically contains two key components: a \emph{converter} and a \emph{generative retriever}. The converter maps each candidate to a discrete identifier
$\mathbf{m}=[m_1,\ldots,m_L]$, where $m_i$ is the $i$-th token and $L$ is the identifier length. The generative retriever is then trained to generate the identifiers of relevant candidates conditioned on the input query. Different frameworks for GUMR can be constructed by varying the converter design,
adopting different token selection strategies during the autoregressive
generation process, and devising alternative loss functions for optimization.
\section{The Proposed Framework}
\label{sec:method}
\begin{figure}[htbp]
  \centering
  \includegraphics[width=1\textwidth]{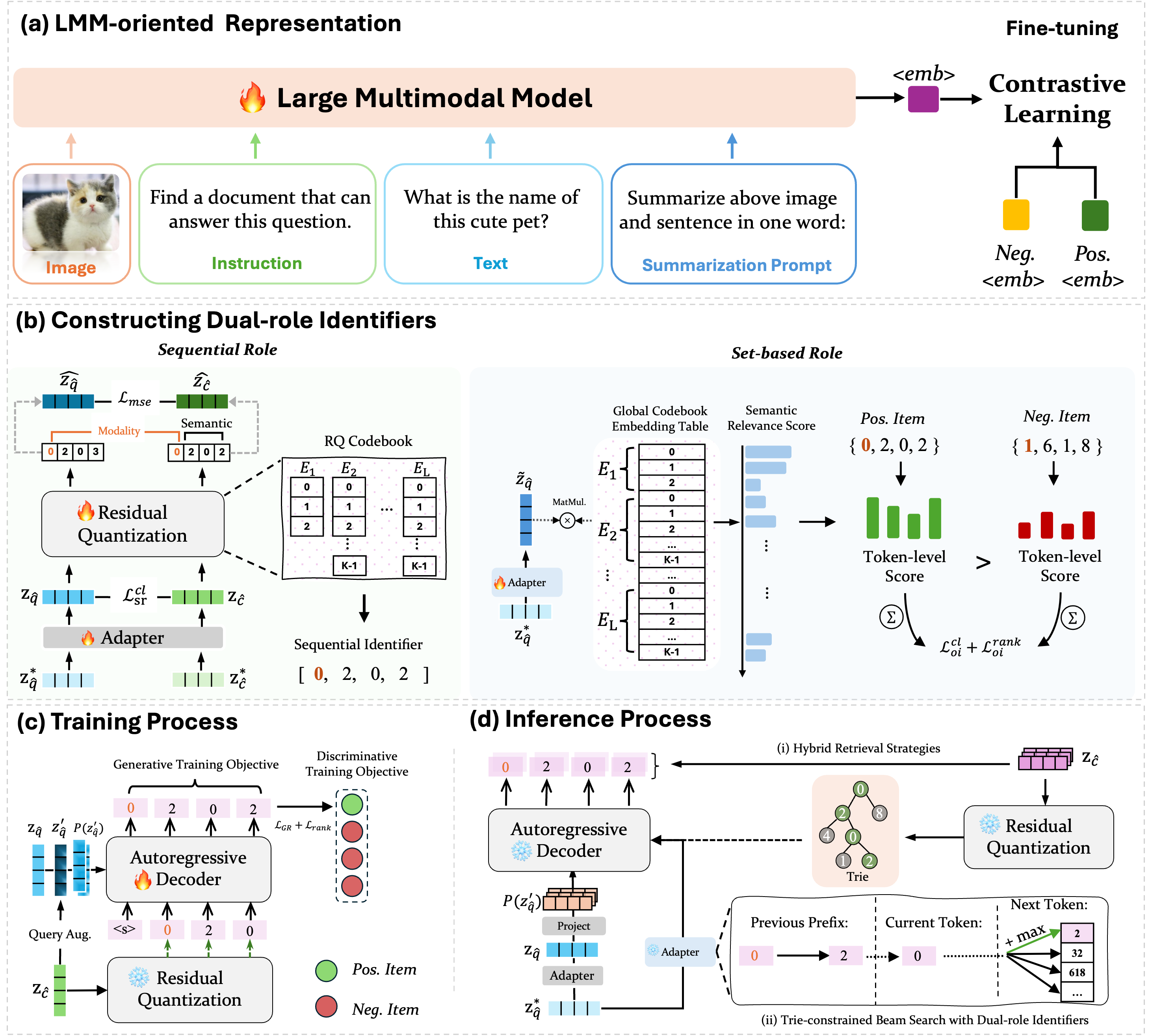}
  \caption{
An overview of the proposed DrIG framework.
}
  \label{fig:framework}
\end{figure}

Figure ~\ref{fig:framework} provides an overview of DrIG, which includes four core stages. (a) \emph{LMM-oriented representation} for encoding queries and candidates into an instruction-aware shared embedding space. (b) \emph{Constructing dual-role identifiers} for converting candidate embeddings into residual-quantized codes that support both sequential generation and set-based relevance estimation. (c) The \emph{training process} for learning the autoregressive decoder with generative and discriminative objectives. (d) The \emph{inference process} for retrieving candidates via dual-guided Trie-constrained beam search and optional dense reranking.  We elaborate these stages in Sections~\ref{sec:lmm_pretrain}--\ref{sec:autoregressive}.

\subsection{Initial Representation in a Shared Embedding Space}
\label{sec:lmm_pretrain}
To cope with the diverse retrieval tasks in universal multimodal retrieval (as shown in Table \ref{tab:mbeir_stats}), we first employ a large multimodal model (LMM) to obtain the initial dense representations of retrieval queries and candidate items by encoding them into a shared embedding space. Then a two-stage fine-tuning strategy is deployed to further improve the capability of the selected LMM for embedding multimodal items in retrieval tasks. The rationale behind this design is twofold. First, benefiting from interleaved vision–language training, LMMs are able to represent multimodal information according to their meanings with prompt. Second, compared with methods that rely on separate encoders for different modalities (e.g., CLIP), LMMs possess stronger language understanding and reasoning capabilities. Consequently, this design not only facilitates similarity estimation between query content and candidate items across different modalities, but also enables accurate comprehension of the search instruction.

\subsubsection{LMM-oriented Representation}
\label{sec:feature extraction}
Inspired by the explicit one-word limitation (EOL) strategy proposed in \cite{jiang2024scaling,jiang2024e5}, we explicitly instruct the adopted LMM to represent multimodal inputs using one word. Following the work by Liu et al. \cite{liu2025lamra}, we append a task-specific summarization prompt to the input as follows: (1) for image-only inputs, the prompt is set to: \textit{<image> Summarize the above image in one word: <emb>}; (2) for text-only inputs, the prompt is set to: \textit{<text> Summarize the above sentence in one word: <emb>}; and (3) for a mixed image-text input, the prompt is set to: \textit{<image><text> Summarize the above image and sentence in one word: <emb>}. In these prompts, \textit{<image>} and \textit{<text>} denote placeholders for the input image and sentence, respectively. Finally, we extract the last hidden state immediately preceding the \textit{<emb>} token as the embedding vector of the input.

\subsubsection{Fine-tuning for Improved Representation}
As demonstrated in prior studies~\cite{liu2025lamra,leenvembed2025,lin2024mm}, most pretrained LMMs are primarily optimized for generative tasks, such as next-token prediction. In contrast, retrieval tasks require distinguishing relevant items from non-relevant ones. Consequently, directly employing an LMM to embed multimodal items for retrieval often leads to inferior performance. After an in-depth comparison of three closely related studies~\cite{liu2025lamra,leenvembed2025,lin2024mm}, we finally follow the two-stage fine-tuning strategy proposed in \cite{liu2025lamra} to further improve the capability of the selected LMM for embedding multimodal items in retrieval tasks. In the first stage, the LMM is adapted for text-to-text retrieval using a natural language inference dataset~\cite{gao-etal-2021-simcse}. In the second stage, we further fine-tune the model on a range of multimodal retrieval tasks based on the M-BEIR dataset (details are provided in Section \ref{sec:dataset}). Across both stages, the training objective is contrastive learning with the InfoNCE loss \cite{oord2018representation}. Due to space limitations, please refer to Section 3.2 of \cite{liu2025lamra} for detailed descriptions of the training procedure.
\subsection{Constructing Dual-role Identifiers}
\label{sec:quantization}
The dense representations of queries and candidate items are continuous and high-dimensional, making them incompatible with autoregressive generation. To bridge this gap, prior generative information retrieval methods convert such representations into discrete identifiers. However, existing approaches typically treat identifiers purely as ordered token sequences, which introduces a critical limitation: the left-to-right generation process may bias the model toward local optima and restrict its ability to capture global similarity. To address this issue, we propose to construct \textit{dual-role identifiers}, which enable complementary interpretations of discrete tokens. Specifically, each identifier simultaneously serves (i) as a sequential representation for autoregressive generation, preserving structural and semantic dependencies, and (ii) as a set-based representation for order-invariant matching, mitigating the limitations of sequential decoding. Based on this design, we construct discrete identifiers for both queries and candidate items through a three-step process, as detailed in the following subsections.
\subsubsection{Sequential Role: Identifier Construction for Autoregressive Generation}
\label{sequential}

We denote a dense representation vector in the shared embedding space described above as $\mathbf{z}^{*}$, and its corresponding  identifier as $\mathbf{m} := (m_1, \ldots, m_L)$. When necessary, we use subscripts such as $\mathbf{z}^{*}_{\widehat{q}}$ and $\mathbf{z}^{*}_{\widehat{c}}$ to distinguish the embeddings of a search query and a candidate item, respectively. Inspired by Kim et al.~\cite{kim2025genius}, the identifier $\mathbf{m}$ is particularly designed such that the first token $m_1$ captures modality distinctions, whereas the remaining tokens $(m_2, \ldots, m_L)$ progressively encode increasingly fine-grained semantic information.

Under the sequential role for autoregressive generation, we adopt a two-stage design for identifier construction. In the first stage, we introduce a lightweight residual MLP module (the \emph{Adapter} block in Figure ~\ref{fig:framework}(b)) to refine the initial dense representation $\mathbf{z}^{*}$ and facilitate the generation of an effective first token $m_1$. Specifically, this module is designed as:
\begin{equation}
\mathbf{z} = \mathbf{z}^{*} + \mathrm{MLP}(\mathbf{z}^{*}),
\label{eq:mlp_adapt}
\end{equation}
where $\mathrm{MLP}(\cdot)$ denotes a learnable multi-layer perceptron, and $\mathbf{z}$ denotes the refined representation vector.

In the second stage, we deploy the residual quantization (RQ)~\cite{lee2022autoregressive,rajput2023recommender} to convert the refined representation $\mathbf{z}$ into a sequential identifier. Let $\mathcal{E} = \{\mathbf{E}_1, \ldots, \mathbf{E}_i, \ldots, \mathbf{E}_L\}$ denote the set of codebooks, and $
\mathbf{E}_i = \left\{ \mathbf{e}_k^{i} \in \mathbb{R}^d \mid k = 1, \dots, |\mathbf{E}_i| \right\}
$ represent the $i$-th codebook, 
where $\mathbf{e}_k^{i}$ is the $k$-th code embedding in $\mathbf{E}_i$, and $|\mathbf{E}_i|$ is the codebook size. For the first codebook $\mathbf{E}_1$, we set its size as $|\mathbf{E}_1| = 3$, corresponding to images, text, and image-text pairs. We initialize the initial residual vector as $\mathbf{r}_0 = \mathbf{z}$ and perform quantization recursively with $L$ steps. At the $i$-th step, we search for the nearest neighbor in the $i$-th codebook $\mathbf{E}_i$ as follows:
\begin{equation}
k_i^{*} = \arg\min_{1 \le k \le |\mathbf{E}_i|}
\left\| \mathbf{r}_{i-1} - \mathbf{e}_k^{i} \right\|^2,
\label{eq:rq_assign}
\end{equation}
where $k_i^{*}$ denotes the selected index at the $i$-th step. Then the residual vector is updated as $\mathbf{r}_i = \mathbf{r}_{i-1} - \mathbf{e}_{k_i^{*}}^{i}$.

Finally, the refined representation vector $\mathbf{z}$ is approximated by summing the selected code embeddings across all quantization codebooks, yielding the quantized vector $\widehat{\mathbf{z}} = \sum_{i=1}^{L} \mathbf{e}_{k_i^{*}}^{i}$. A key property of residual quantization is that the code embedding selected at each level represents the residual information specific to that level, thus the information is progressively separated across quantization levels.

To ensure the quantization quality, our training objective incorporates three types of loss functions that are computed over the same mini-batch of query–candidate pairs, denoted as $B_{sr}=\{(\mathbf{z}_{\widehat{q}}^{\,i}, \mathbf{z}_{\widehat{c}}^{\,i})|1\leq i\leq|B_{sr}|\}$, where $(\mathbf{z}_{\widehat{q}}^{\,i}, \mathbf{z}_{\widehat{c}}^{\,i})$ denotes the $i$-th pair, and $|B_{sr}|$ denotes the corresponding batch size. The first is the residual quantization loss defined in Equation-\ref{eq:rq_loss},
\begin{equation}
\mathcal{L}_{\mathrm{rq}}
=\frac{1}{2|B_{sr}|}
\sum_{\mathbf{z}\in B_{sr}}^{}
\sum_{i=1}^{L}
\left\|
\mathbf{r}_{i-1} - \mathrm{sg}\!\left(\mathbf{e}_{k_i^{*}}^{i}\right)
\right\|^{2},
\label{eq:rq_loss}
\end{equation}
Here, omitting the subscripts of $\mathbf{z}$ indicates that $\mathbf{z}$ may represent either a search query or a candidate item. This also explains why $2|B_{sr}|$ is used as the denominator when computing the mean value. This loss encourages each residual vector to align with its assigned code embedding so that the quantization error is progressively reduced across quantization levels, where $\mathrm{sg}(\cdot)$ denotes the stop-gradient operator, preventing gradients from directly updating the code embeddings. Instead, the code embeddings are updated via an exponential moving average (EMA)~\cite{razavi2019generating} over training steps to ensure stable updates.

The second loss function is the contrastive loss defined in Equation-\ref{eq:contrastive_loss},
\begin{equation}
\mathcal{L}_{\mathrm{sr}}^{\mathrm{cl}}
=
-\frac{1}{|B_{sr}|}
\sum_{i=1}^{|B_{sr}|}
\log
\frac{
\exp\!\left(
\cos(\mathbf{z}_{\widehat{q}}^{\,i}, \mathbf{z}_{\widehat{c}}^{\,i}) / \tau
\right)
}{
\sum_{j=1}^{|B_{sr}|}
\exp\!\left(
\cos(\mathbf{z}_{\widehat{q}}^{\,i}, \mathbf{z}_{\widehat{c}}^{\,j}) / \tau
\right)
},
\label{eq:contrastive_loss}
\end{equation}
For the $i$-th query $\mathbf{z}_{\widehat{q}}^{\,i}$ in the mini-batch, we view $\mathbf{z}_{\widehat{c}}^{\,i}$ as the positive candidate, whereas the candidates paired with the remaining queries are treated as in-batch negative candidates. Since the query is guided by an instruction that specifies the desired target type, the pre-quantization embedding space should be structured so that each query is aligned with targets consistent with both its semantics and retrieval intent. This loss function encourages matched query--target pairs to be close in the embedding space while pushing apart mismatched candidates, thereby inducing modality-level grouping and semantic alignment within each group.

The third loss function is the MSE loss defined in Equation-\ref{eq:mse_loss},
\begin{equation}
\mathcal{L}_{\mathrm{mse}}
=
\frac{1}{|B_{sr}|}
\sum_{i=1}^{|B_{sr}|}
\left\|
\widehat{\mathbf{z}}_{\widehat{q}}^{\,i}
-
\widehat{\mathbf{z}}_{\widehat{c}}^{\,i}
\right\|_2^2,
\label{eq:mse_loss}
\end{equation}
where $\widehat{\mathbf{z}}_{\widehat{q}}^{\,i}$ and $\widehat{\mathbf{z}}_{\widehat{c}}^{\,i}$ denote the quantized representations obtained from $\mathbf{z}_{\widehat{q}}^{\,i}$ and $\mathbf{z}_{\widehat{c}}^{\,i}$, respectively. This loss function further preserves pairwise consistency after quantization by keeping the quantized query representation close to its matched target candidate representation.

Finally, the overall objective for optimizing the identifier’s sequential role is formulated by combining the above three loss functions as
\begin{equation}
\mathcal{L}_{\mathrm{sr}}
=
\mathcal{L}_{\mathrm{sr}}^{\mathrm{cl}}
+
\beta_{1}\mathcal{L}_{\mathrm{rq}}
+
\beta_{2}\mathcal{L}_{\mathrm{mse}},
\label{eq:combined_loss}
\end{equation}
Empirically, $\mathcal{L}_{\mathrm{sr}}^{\mathrm{cl}}$ is typically on the order of $10^{-1}$, while $\mathcal{L}_{\mathrm{rq}}$ and $\mathcal{L}_{\mathrm{mse}}$ are on the order of $10^{-2}$. To well balance their contributions, we introduce the weighting coefficients $\beta_{1}$ and $\beta_{2}$. 

\subsubsection{Set-based Role: Identifier Construction for Order-invariant Matching}
\label{order_invariant}

Under the set-based role for order-invariant matching, we aim to mitigate the limitations introduced by prefix-dependent autoregressive decoding. While the sequential identifier $\mathbf{m}$ enables token-by-token generation, inference in generative retrieval typically relies on constrained beam search, where identifiers are expanded based on prefix-level likelihoods. Consequently, once the prefix of a relevant identifier is pruned at an early decoding step, the corresponding target becomes irrecoverable. This prefix-sensitive behavior often leads to local optima, particularly when multiple candidates share similar prefixes.

To address this issue, a key desideratum is to provide a global, prefix-independent relevance signal. Rather than introducing a separate identifier space, we reinterpret the sequential identifier $\mathbf{m} = [m_1, \ldots, m_L]$ as an unordered set. This set-based reinterpretation discards positional information while preserving the underlying semantic tokens, thereby enabling order-invariant matching. To operationalize identifiers under the set-based role, we adopt a two-step design. 

Noting that identifiers consist of discrete code tokens selected from RQ codebooks, the first step aims to map the query representation into the same token space, thereby facilitating direct token-level relevance estimation. Specifically, given a query embedding $\mathbf{z}_{\widehat{q}}$, we refine it via an MLP module as $\tilde{\mathbf{z}}_{\widehat{q}} = \mathrm{MLP}_{\mathrm{oi}}(\mathbf{z}^{*}_{\widehat{q}})$. Based on the refined representation, we compute a token-level score vector as
\begin{equation}
\mathbf{s}_{\widehat{q}}
=
\log\!\bigl(1 + \mathrm{ReLU}(\tilde{\mathbf{z}}_{\widehat{q}} \bar{\mathbf{E}}^\top)\bigr)
\in \mathbb{R}^{V_T},
\end{equation}
where $\bar{\mathbf{E}} \in \mathbb{R}^{V_T \times d}$ denotes the fixed global codebook embedding table, constructed by aggregating code embeddings from all frozen RQ codebooks in $\mathcal{E}$, and $V_T$ is the size of the resulting code-token vocabulary. In particular, we interpret each dimension of $\mathbf{s}_{\widehat{q}}$ as a relevance indicator measuring how well the semantic content encoded by the corresponding code token aligns with the query.

Given this vocabulary-wise score vector and a candidate identifier $\mathbf{m} = \{m_1, \ldots, m_L\}$, the second step computes an order-invariant relevance score by aggregating the token-level scores associated with the tokens in $\mathbf{m}$:
\begin{equation}
\label{soi}
s_{\mathrm{oi}}(\mathbf{z}_{\widehat{q}}, \mathbf{m})
=
\sum_{l=1}^{L}
\mathbf{s}_{\widehat{q}}[\psi(l, m_l)],
\end{equation}
\noindent where the level-specific mapping function $\psi(l, m_l) = m_l + \sum_{i=1}^{l-1} |\mathbf{E}_i|$ maps the level-wise token index $m_l$ to its global vocabulary index, and $[\,\cdot\,]$ denotes the corresponding indexing operation. 

Finally, this formulation yields a set-level relevance score between the query and the candidate that is invariant to token order and independent of the decoding prefix.

To optimize the identifier’s set-based role for order-invariant matching, we adopt two loss functions computed over a mini-batch of query-candidate pairs, denoted as $B_{\mathrm{oi}}
=
\{(\mathbf{z}_{\widehat{q}}^{\,i}, \mathbf{z}_{\widehat{c}}^{\,i}) \mid 1 \le i \le |B_{\mathrm{oi}}|\},
$
where $|B_{\mathrm{oi}}|$ denotes the corresponding batch size. The first loss function is a contrastive objective, 
\begin{equation}
\mathcal{L}_{\mathrm{oi}}^{\mathrm{cl}}
=
-\frac{1}{|B_{\mathrm{oi}}|}
\sum_{i=1}^{|B_{\mathrm{oi}}|}
\log
\frac{
\exp\!\left(
s_{\mathrm{oi}}(\mathbf{z}_{\widehat{q}}^{\,i}, \mathbf{m}_i) / \tau
\right)
}{
\sum_{j=1}^{|B_{\mathrm{oi}}|}
\exp\!\left(
s_{\mathrm{oi}}(\mathbf{z}_{\widehat{q}}^{\,i}, \mathbf{m}_j) / \tau
\right)
}.
\end{equation}
\noindent Analogous to Equation-\ref{eq:contrastive_loss}, the key difference lies in replacing the query–candidate relevance score with the order-invariant score defined in Equation-\ref{soi}.

The second loss function is based on the margin ranking loss. To enhance discrimination against hard negatives, we identify the most challenging in-batch negative for each query according to dense vector-based cosine similarity: $j_i^{-} = \arg\max_{j \neq i}
\cos(\mathbf{z}_{\widehat{q}}^{\,i}, \mathbf{z}_{\widehat{c}}^{\,j})$. The corresponding identifier $\mathbf{m}_{j_i^{-}}$  is used as the hard negative for the $i$-th query. Then a fixed margin $\rho$ is enforced between the positive and hard-negative scores:
\begin{equation}
\mathcal{L}_{\mathrm{oi}}^{\mathrm{rank}}
=
\frac{1}{|B_{\mathrm{oi}}|}
\sum_{i=1}^{|B_{\mathrm{oi}}|}
\max\!\Bigl(
0,\,
\rho
-
\bigl(
s_{\mathrm{oi}}(\mathbf{z}_{\widehat{q}}^{\,i}, \mathbf{m}_i)
-
s_{\mathrm{oi}}(\mathbf{z}_{\widehat{q}}^{\,i}, \mathbf{m}_{j_i^{-}})
\bigr)
\Bigr).
\end{equation}
Finally, the overall objective for optimizing the identifier’s set-based role is defined as the combination of the two losses: $\mathcal{L}_{\mathrm{oi}}
=
\mathcal{L}_{\mathrm{oi}}^{\mathrm{cl}}
+
\mathcal{L}_{\mathrm{oi}}^{\mathrm{rank}}$.
Unlike the losses in the sequential role, the two objectives in the set-based role are empirically of comparable magnitude, and we therefore assign them equal importance.
\subsection{Generative Retrieval with Dual-role Identifiers}
\label{sec:autoregressive}
To fully exploit the complementary properties of dual-role identifiers, we use them in two ways: the sequential role supports autoregressive decoding, while the set-based role provides a global relevance prior during search. We first describe how the two roles are combined at inference time, and then introduce the corresponding training objectives.

\subsubsection{Inference Process}
\label{sec:inference}
During inference, we exploit the two complementary roles of the same identifier to guide the generation process. Specifically, the sequential role provides fine-grained and order-sensitive token predictions through autoregressive decoding, while the set-based role provides a global, prefix-independent relevance prior. By combining sequential decoding scores with order-invariant global relevance signals, our method alleviates the local-optimum issue in constrained beam search.

To condition the decoder on the query representation, we employ a lightweight projector that maps the query embedding into $N$ prefix embeddings:
\begin{equation}
\mathbf{P}(\mathbf{z}_{\widehat q})
=
\mathrm{Reshape}\!\left(\mathrm{MLP}(\mathbf{z}_{\widehat q})\right)
\in
\mathbb{R}^{N \times d'},
\label{eq:prefix_proj}
\end{equation}
where $d'$ denotes the hidden dimension of the decoder. Based on these prefix embeddings, decoding proceeds over the sequential identifier space using constrained beam search.
More specifically, we build a Trie (prefix tree) from the sequential identifiers of all candidates and use it to constrain generation. A newly expanded prefix is considered valid only if it matches at least one candidate identifier prefix. We express this constraint by the validation function
\begin{equation}
\delta(\mathbf{t}_{\le i})=
\begin{cases}
0, & \text{if } [t_1,\ldots,t_i] \text{ is a valid prefix},\\
-\infty, & \text{otherwise},
\end{cases}
\end{equation}
where $\mathbf{t}_{\le i}=[t_1,\ldots,t_i]$ denotes the prefix after expanding the current token.

With the prefix constraint in place, we describe how the sequential decoding
score is computed at each step. Given the query embedding
$\mathbf{z}_{\widehat q}$ and the current prefix $\mathbf{t}_{<i}$, the decoder
produces the hidden state for the $i$-th decoding step as
\begin{equation}
\mathbf{h}_i
=
\mathrm{Decoder}(\mathbf{P}(\mathbf{z}_{\widehat q}), \mathbf{t}_{<i})
\in \mathbb{R}^{d'}.
\end{equation}
Let $\mathbf{E}^{\mathrm{dec}}_i \in \mathbb{R}^{V_i \times d'}$ denote the
output embedding table for code tokens at the $i$-th identifier level, where
$V_i$ is the corresponding vocabulary size. To preserve the semantics encoded
in the quantized space, we initialize the embeddings using the
corresponding RQ codebook vectors projected into the decoder embedding space.
The unnormalized autoregressive score for token $t_i$ is then
$\mathbf{E}^{\mathrm{dec}}_i[t_i]\cdot \mathbf{h}_i$, and the full sequential
decoding score for a generated identifier $\mathbf{t}=[t_1,\ldots,t_L]$ is
\begin{equation}
\eta(\mathbf{t};\mathbf{z}_{\widehat q})
=
\sum_{i=1}^{L}
\mathbf{E}^{\mathrm{dec}}_i[t_i]\cdot \mathbf{h}_i,
\end{equation}
which accumulates order-sensitive evidence along the autoregressive decoding
path.

To incorporate the set-based role, we further introduce a global relevance
prior derived from the order-invariant scoring function
$s_{\mathrm{oi}}(\cdot,\cdot)$ defined in Section~\ref{order_invariant}.
For a partial prefix $\mathbf{t}_{\le i}$, we consider all candidate
identifiers whose sequential forms share this prefix and compute
\begin{equation}
\label{GlobalPrior}
\phi(\mathbf{t}_{\le i};\mathbf{z}_{\widehat q})
=
\max_{\mathbf{m}\in \mathcal{C}_{\mathbf{t}_{\le i}}}
s_{\mathrm{oi}}(\mathbf{z}_{\widehat q}, \mathbf{m}),
\end{equation}
where $\mathcal{C}_{\mathbf{t}_{\le i}}$ denotes the set of candidate
identifiers whose sequential forms share the prefix $\mathbf{t}_{\le i}$.
Each identifier in this set is evaluated by $s_{\mathrm{oi}}$ under its
set-based interpretation. Thus, $\phi(\mathbf{t}_{\le i};\mathbf{z}_{\widehat q})$
provides the strongest global relevance signal among candidates reachable
from the current prefix.
We then combine the prefix validity constraint, the sequential decoding score,
and the set-based global prior into a unified expansion score:
\begin{equation}
\label{Eq_g_obj}
f(\mathbf{t}_{\le i};\mathbf{z}_{\widehat q})
=
\delta(\mathbf{t}_{\le i})
+
\eta(\mathbf{t}_{<i};\mathbf{z}_{\widehat q})
+
\mathbf{E}^{\mathrm{dec}}_{i}[t_i]\cdot\mathbf{h}_{i}
+
\lambda\phi(\mathbf{t}_{\le i};\mathbf{z}_{\widehat q}).
\end{equation}
Here, $\lambda$ is a weighting hyperparameter that balances the contribution
of the set-based global prior against the sequential decoding score.
Each term in Equation-\ref{Eq_g_obj} serves a distinct but complementary purpose. The validation score
$\delta(\mathbf{t}_{\le i})$ ensures that the newly expanded prefix remains
consistent with at least one valid candidate identifier. The sequential
component
$\eta(\mathbf{t}_{<i};\mathbf{z}_{\widehat q})
+\mathbf{E}^{\mathrm{dec}}_{i}[t_i]\cdot\mathbf{h}_{i}$
provides fine-grained, order-sensitive token-level evidence from the
autoregressive decoder, but may still be vulnerable to local optima caused by
early pruning. In contrast, the set-based prior
$\phi(\mathbf{t}_{\le i};\mathbf{z}_{\widehat q})$ supplies a global relevance
signal that is independent of the local decoding likelihood. By combining the
two roles of the same identifier in this way, the decoder can maintain both
local generation fidelity and global retrieval awareness throughout the search
process.

\subsubsection{Query Augmentation via Interpolation}
In multimodal generative retrieval, the limited representation capacity of discrete identifiers compared with continuous embeddings poses a challenge for model generalization, especially when training data is limited. To alleviate this issue, we introduce a query augmentation strategy based on query--target interpolation in the continuous embedding space.
Concretely, we construct an augmented query embedding by interpolating the query representation $\mathbf{z}_{\widehat q}$ and the representation of its paired target $\mathbf{z}_{\widehat c}$:
\begin{equation}
\mathbf{z}'_{\widehat q}
=
\mu \cdot \mathbf{z}_{\widehat q}
+
(1-\mu) \cdot \mathbf{z}_{\widehat c},
\label{eq:query_interp}
\end{equation}
where $\mu$ is randomly sampled from a Beta distribution, i.e., $\mu \sim \mathrm{Beta}(\alpha,\alpha)$. This strategy generates diverse augmented queries while preserving semantic relevance to the target, thereby improving robustness to query variation.

\subsubsection{Generative Training Objective}
During training, the decoder is conditioned on the augmented query embedding through the projector $\mathbf{P}(\mathbf{z}'_{\widehat q})$. The generative loss is defined as the token-level cross-entropy over the target sequential identifier of the paired candidate:
\begin{equation}
\mathcal{L}_{\mathrm{GR}}\!\left(\mathbf{P}(\mathbf{z}'_{\widehat{q}}), \mathbf{m}_{\widehat{c}}\right)
=
-\sum_{k=1}^{L}
\log p\!\left(
m^{\widehat{c}}_k
\mid
\mathbf{P}(\mathbf{z}'_{\widehat{q}}),\,
m^{\widehat{c}}_{<k}
\right).
\label{eq:gr_loss}
\end{equation}
This objective encourages the decoder to generate the target identifier sequence conditioned on the augmented query, thereby learning a robust mapping from continuous query semantics to discrete retrieval identifiers.

\subsubsection{Discriminative Training Objective}
While generative learning focuses on predicting one correct identifier sequence, retrieval ultimately requires ranking candidates according to relevance. To reduce this discrepancy, we introduce a discriminative ranking objective that encourages the decoder to assign higher sequence-level scores to relevant target candidates than to hard negatives.

Similar to the hard-negative selection used in the set-based role, we construct the discriminative objective over a mini-batch of query--candidate pairs, denoted as
$B_{\mathrm{dec}}
=
\{(\mathbf{z}_{\widehat{q}}^{\,i}, \mathbf{z}_{\widehat{c}}^{\,i}, \mathbf{m}_{\widehat{c}}^{\,i}) \mid 1 \le i \le |B_{\mathrm{dec}}|\}$,
where $\mathbf{m}_{\widehat{c}}^{\,i}$ is the sequential identifier of the paired target candidate. For each query $i$, we retain $\mathbf{m}_{\widehat{c}}^{\,i}$ as the positive identifier and select the most similar non-matching candidate in the batch according to cosine similarity, i.e.,
$j_i^{-} = \arg\max_{j \neq i}
\cos(\mathbf{z}_{\widehat{q}}^{\,i}, \mathbf{z}_{\widehat{c}}^{\,j})$.
The sequential identifier of the selected candidate,
$\mathbf{m}_{\widehat{c}}^{\,j_i^{-}}$, is then treated as the hard negative for query $i$.
To compare candidates, we score each candidate sequential identifier under teacher forcing by averaging the token-level log-likelihoods assigned by the decoder. Given a candidate sequential identifier $\mathbf{m}=(m_1,\ldots,m_L)$, we define its sequence-level score as
\begin{equation}
s_{\mathrm{seq}}(\mathbf{P}(\mathbf{z}_{\widehat{q}}^{\prime\,i}), \mathbf{m})
=
\frac{1}{L}
\sum_{k=1}^{L}
\log p\!\left(
m_k
\mid
\mathbf{P}(\mathbf{z}_{\widehat{q}}^{\prime\,i}),\,
m_{<k}
\right).
\label{eq:seq_score}
\end{equation}
Based on this scoring function, the positive and negative sequence-level scores are obtained as
$s_i^{+}
=
s_{\mathrm{seq}}(\mathbf{P}(\mathbf{z}_{\widehat{q}}^{\prime\,i}), \mathbf{m}_{\widehat{c}}^{\,i})$
and
$s_i^{-}
=
s_{\mathrm{seq}}(\mathbf{P}(\mathbf{z}_{\widehat{q}}^{\prime\,i}), \mathbf{m}_{\widehat{c}}^{\,j_i^{-}})$,
respectively.
We adopt an adaptive margin derived from the teacher similarity signal in the original embedding space:
\begin{equation}
\mathcal{M}_i
=
\max\!\left(
0,\,
\bigl(
\cos(\mathbf{z}_{\widehat{q}}^{\,i}, \mathbf{z}_{\widehat{c}}^{\,i})
-
\cos(\mathbf{z}_{\widehat{q}}^{\,i}, \mathbf{z}_{\widehat{c}}^{\,j_i^{-}})
\bigr)\cdot \gamma
\right),
\label{eq:adaptive_margin}
\end{equation}
where $\gamma$ controls the scale of the margin. This design yields a softer constraint when the positive and negative candidates are semantically close, and a stronger one when they are well separated in the teacher embedding space.
We then optimize a smooth pairwise ranking loss:
\begin{equation}
\label{rankloss}
\mathcal{L}_{\mathrm{rank}}
=
\frac{1}{|B_{\mathrm{dec}}|}
\sum_{i=1}^{|B_{\mathrm{dec}}|}
\log\!\left(
1 + \exp\bigl(\mathcal{M}_i - (s_i^{+} - s_i^{-})\bigr)
\right).
\end{equation}

The final decoder objective combines the generative loss and the discriminative
ranking loss as
$
\mathcal{L}_{\mathrm{decoder}}
=
\mathcal{L}_{\mathrm{GR}}
+
\mathcal{L}_{\mathrm{rank}}.
$
Here, $\mathcal{L}_{\mathrm{GR}}$ provides token-level supervision for generating
the correct identifier sequence, while $\mathcal{L}_{\mathrm{rank}}$ further
encourages the decoder to assign higher sequence-level likelihood to relevant
identifiers than to hard negatives. Compared with using only the generative
loss, this combined objective better aligns autoregressive identifier generation
with the retrieval objective, where semantically similar distractors should be
ranked below the paired target.

\subsection{Hybrid Retrieval Strategies}
A closer examination of the previous steps in Sections~\ref{sec:quantization} and \ref{sec:autoregressive} reveals that discrete quantization is a crucial component of generative retrieval, which inevitably incurs information loss due to the approximation of continuous embeddings by finite codebooks. Although residual quantization preserves a coarse-to-fine semantic structure and enables efficient identifier generation, subtle distinctions between highly similar candidates may not be fully captured in the discrete space. To mitigate this limitation, we adopt a hybrid retrieval strategy that combines generative retrieval with dense vector-based reranking. Specifically, we first perform generative retrieval under dual guidance to obtain a ranked list of candidate identifiers and select the top-$k$ candidates. These candidates are then reranked using the cosine similarity between the query embedding $\mathbf{z}_q$ and candidate embeddings $\mathbf{z}_c$.

From a computational perspective, generative retrieval avoids exhaustive similarity computation over the entire candidate pool, whose complexity would be $\mathcal{O}(|\mathcal{C}| d)$ for dense retrieval, where $|\mathcal{C}|$ denotes the number of candidates and $d$ the embedding dimension. Instead, the decoder explores only valid identifier prefixes via constrained beam search, significantly reducing the effective search space. The additional reranking stage operates only on the top-$k$ candidates, incurring a cost of $\mathcal{O}(k d)$ with $k \ll |\mathcal{C}|$, making the overhead negligible in practice. This hybrid design therefore preserves the scalability advantage of generative retrieval while compensating for quantization-induced information loss through continuous fine-grained scoring.

\section{Experiments}
\label{sec:experiments}
In this work, we investigate the following research questions: (1) \textbf{RQ1}: How does DrIG perform compared to existing dense vector-based approaches and generative retrieval methods in the context of universal multimodal retrieval? (2) \textbf{RQ2}: To what extent can DrIG generalize to standard text-to-image retrieval benchmarks under both zero-shot and in-domain training settings? (3) \textbf{RQ3}: How does each key component of DrIG contribute to its overall retrieval performance? (4) \textbf{RQ4}: How does DrIG balance retrieval effectiveness and inference efficiency relative to baseline dense vector-based and generative retrieval methods?
Beyond these primary research questions, we further analyze the impact of key hyperparameters and conduct qualitative case studies to better understand the behaviors and capabilities of DrIG.

\subsection{Experimental Setting}
\subsubsection{Datasets}
\label{sec:dataset}

%%%%M-BEIR dataset
\begin{table}[t]
\caption{An overview of M-BEIR. The columns refer to the task-specific retrieval setup, the constituent datasets, the involved domains, the number of queries across the Train/Validation/Test splits (\#Query), the average number of relevant items per query (\#Rel./Query), and the total number of candidate items (\#Candid.).}
\label{tab:mbeir_stats}
\centering
\resizebox{\columnwidth}{!}{%
\begin{tabular}{llcccccccc}
\toprule
\multirow{2}{*}{\textbf{Task}} &
\multirow{2}{*}{\textbf{Dataset}} &
\multirow{2}{*}{\textbf{Domain}} &
\multicolumn{3}{c}{\textbf{\#Query}} &
\multicolumn{3}{c}{\textbf{\#Rel./Query}} &
\multirow{2}{*}{\textbf{\#Candid.}} \\
\cmidrule(lr){4-6}\cmidrule(lr){7-9}
& & & \textbf{Train} & \textbf{Val.} & \textbf{Test} & \textbf{Train} & \textbf{Val.} & \textbf{Test} & \\
\midrule

1. $q_t \rightarrow c_i$
& VisualNews~\cite{liu2021visualnewsbenchmarkchallenges} & News
& 99K & 20K & 20K
& 1.0 & 1.0 & 1.0
& 542K \\
& MSCOCO~\cite{lin2015microsoftcococommonobjects} & Misc.
& 100K & 24.8K & 24.8K
& 1.0 & 1.0 & 1.0
& 5K \\
& Fashion200K~\cite{han2017automaticspatiallyawarefashionconcept} & Fashion
& 15K & 1.7K & 1.7K
& 3.3 & 3.1 & 2.8
& 201K \\
\midrule

2. $q_t \rightarrow c_t$
& WebQA~\cite{chang2022webqamultihopmultimodalqa} & Wiki
& 16K & 1.7K & 2.4K
& 2.0 & 2.0 & 2.0
& 544K \\
\midrule

3. $q_t \rightarrow (c_i,c_t)$
& EDIS~\cite{liu2023edisentitydrivenimagesearch} & News
& 26K & 3.2K & 3.2K
& 2.6 & 2.6 & 2.6
& 1M \\
& WebQA~\cite{chang2022webqamultihopmultimodalqa} & Wiki
& 17K & 1.7K & 2.5K
& 1.4 & 1.4 & 1.4
& 403K \\
\midrule

4. $q_i \rightarrow c_t$
& VisualNews~\cite{liu2021visualnewsbenchmarkchallenges} & News
& 100K & 20K & 20K
& 1.0 & 1.0 & 1.0
& 537K \\
& MSCOCO~\cite{lin2015microsoftcococommonobjects} & Misc.
& 113K & 5K & 5K
& 5.0 & 5.0 & 5.0
& 25K \\
& Fashion200K~\cite{han2017automaticspatiallyawarefashionconcept} & Fashion
& 15K & 4.8K & 4.8K
& 1.0 & 1.0 & 1.0
& 61K \\
\midrule

5. $q_i \rightarrow c_i$
& NIGHTS~\cite{fu2023dreamsimlearningnewdimensions} & Misc.
& 16K & 2K & 2K
& 1.0 & 1.0 & 1.0
& 40K \\
\midrule

6. $(q_i,q_t)\rightarrow c_t$
& OVEN~\cite{hu2023opendomainvisualentityrecognition} & Wiki
& 150K & 50K & 50K
& 8.5 & 10.0 & 9.9
& 676K \\
& InfoSeek~\cite{chen2023pretrainedvisionlanguagemodels} & Wiki
& 141K & 11K & 11K
& 6.8 & 6.7 & 6.5
& 611K \\
\midrule

7. $(q_i,q_t)\rightarrow c_i$
& FashionIQ~\cite{wu2020fashioniqnewdataset} & Fashion
& 16K & 2K & 6K
& 1.0 & 1.0 & 1.0
& 74K \\
& CIRR~\cite{liu2021imageretrievalreallifeimages} & Misc.
& 26K & 2K & 4K
& 1.0 & 1.0 & 1.0
& 21K \\
\midrule

8. $(q_i,q_t)\rightarrow (c_i,c_t)$
& OVEN~\cite{hu2023opendomainvisualentityrecognition} & Wiki
& 157K & 14.7K & 14.7K
& 17.8 & 17.5 & 17.7
& 335K \\
& InfoSeek~\cite{chen2023pretrainedvisionlanguagemodels} & Wiki
& 143K & 17.6K & 17.6K
& 9.1 & 7.5 & 7.5
& 481K \\
\midrule

\multicolumn{2}{c}{\textbf{M-BEIR}~\cite{wei2024uniir}}
& \multicolumn{1}{c}{4 domains}
& 1.1M & 182K & 190K
& 6.5 & 5.9 & 5.7
& 5.6M \\
\bottomrule
\end{tabular}
}
\end{table}

To evaluate universal multimodal retrieval, we adopt the widely recognized M-BEIR benchmark~\cite{wei2024uniir}. M-BEIR integrates 10 datasets across 4 domains and encompasses 8 retrieval task types, where both queries and candidates can be text, images, or image-text pairs. Table \ref{tab:mbeir_stats} summarizes the key statistics of these datasets, including task type, source, domain, query splits, average relevant candidates per query, and total candidate pool size. We refer readers to the original paper~\cite{wei2024uniir} for detailed descriptions of the constituent datasets and their preprocessing procedures. Following previous studies~\cite{wei2024uniir,liu2025lamra}, our evaluation on M-BEIR is conducted under two distinct settings: (i) \textit{local-pool retrieval}, where each query is searched only against the candidate pool of its corresponding dataset; and (ii) \textit{global-pool retrieval}, where candidates from all datasets are combined into a single unified pool.

In addition, we include Flickr30K~\cite{young2014image} and MSCOCO (the latter being a subset of M-BEIR) to facilitate additional comparisons with representative generative text-to-image retrieval methods.

\subsubsection{Evaluation Metric}
We evaluate retrieval effectiveness using query-level mean Recall@K. 
\begin{equation}
\mathrm{Recall@K}
=
\frac{1}{|\mathcal{Q}|}
\sum_{q \in \mathcal{Q}}
\mathbb{I}
\left(
\mathrm{TopK}(q) \cap \mathcal{G}_q \neq \emptyset
\right),
\end{equation}
where $\mathcal{Q}$ denotes the query set, $\mathcal{G}_q$ is the set of 
ground-truth candidate items for query $q$, and $\mathrm{TopK}(q)$ denotes the 
set of top-$K$ retrieved candidate items. $\mathbb{I}(\cdot)$ is the indicator function. Since a query may have multiple relevant candidates, we define a successful retrieval as the presence of at least one relevant candidate within the top-$K$ retrieved items. Following prior studies, we evaluate the performance in terms of Recall@1, 
Recall@5, and Recall@10. Due to space constraints, Recall@k is abbreviated as R@k in some tables.
\subsubsection{Implementation Details}

\begin{table}[t]
\centering
\small
\setlength{\tabcolsep}{4pt}
\renewcommand{\arraystretch}{1}
\caption{The key hyperparameters and implementation settings.}
\label{tab:hyperparameters}
\begin{tabular}{ll @{\hspace{2em}} ll}
\toprule
\textbf{Parameter} & \textbf{Value} & \textbf{Parameter} & \textbf{Value} \\
\midrule
Random seed & 2026 
& Modality codebook size & 3 \\
\midrule
\multicolumn{4}{c}{\textit{Identifier Construction}} \\
\midrule
Codebook vocabulary size & $\{256,1024,2048,4096\}$ 
& RQ levels & $\{4,6,8\}$ \\
Training epochs / batch size & 20 / 2000
& Learning rate & $1\times10^{-4}$ \\
\midrule
\multicolumn{4}{c}{\textit{Set-based and Sequential Roles}} \\
\midrule
Set-based epochs / batch size & 10 / 512
& Set-based learning rate & $2\times10^{-4}$ \\
Sequential epochs / batch size & 30 / 256
& Sequential learning rate & $1\times10^{-4}$ \\
Warmup steps & 500
& & \\
\midrule
\multicolumn{4}{c}{\textit{Evaluation}} \\
\midrule
Beam size & $\{1,5,10,20,30,40,50\}$
& Guidance scale $\lambda$ & $\{0.0,0.1,\ldots,1.0\}$ \\
\bottomrule
\end{tabular}
\end{table}

Following LamRA~\cite{liu2025lamra}, we adopt Qwen2-VL~\cite{wang2024qwen2} as the multimodal encoder to produce 3584-dimensional embeddings for both queries and candidates. The generative retriever uses T5-small~\cite{raffel2020exploring} as the autoregressive decoder. For identifier construction, we use modality-aware residual quantization. The default setting prepends a first-level 
modality codebook of size 3 to distinguish image, text, and image--text candidates, followed by residual codebooks for semantic encoding. The codebooks are initialized by k-means and updated with EMA~\cite{razavi2019generating}  during training.  
DrIG is trained in three stages: 
(i) RQ-based identifier construction using Equation-\ref{eq:combined_loss}; 
(ii) set-based role optimization for global, prefix-independent relevance 
guidance; and 
(iii) sequential identifier generation by training the prefix projector and T5 decoder. 
The RQ module is trained for 20 epochs using AdamW with batch size 2000, learning rate $1\times10^{-4}$, weight decay 0.01, loss weights $\beta_1=\beta_2=100.0$, and contrastive 
temperature $\tau=0.01$.
The set-based module is a two-layer MLP with hidden size 4096 and is trained for 10 epochs using AdamW with batch size 512, learning rate $2\times10^{-4}$, weight decay $1\times10^{-4}$, margin $\rho=0.1$, and temperature $\tau=0.01$. The sequential generation stage uses AdamW with batch size 256, learning rate $1\times10^{-4}$, weight decay $1\times10^{-4}$, 500 warmup steps, maximum gradient norm 1.0, and query interpolation coefficient 
$\alpha=2.0$.
During inference, we perform Trie-constrained beam search over valid candidate identifiers. Unless otherwise specified, the beam size and guidance scale are set to 50 and 1.0, respectively. The generated top-$k$ candidates can be further reranked using dense vector-based similarity. All experiments are conducted with mixed precision and distributed data parallel training on four NVIDIA A100 40GB GPUs. The main hyperparameters are summarized in Table~\ref{tab:hyperparameters}.

\subsubsection{Baseline Methods}
In this work, we compare our method against two distinct groups of baselines. The first group consists of \textit{dense vector-based methods}, including CLIP~\cite{radford2021learning}, BLIP~\cite{li2022blipbootstrappinglanguageimagepretraining}, their respective variants proposed in~\cite{wei2024uniir} (CLIP-SF, CLIP-FF, BLIP-SF, and BLIP-FF), and LamRA~\cite{liu2025lamra}. Notably, LamRA represents the current state-of-the-art approach. The second group comprises \textit{generative retrieval methods}. For universal multimodal retrieval, we consider GENIUS~\cite{kim2025genius} as the current state-of-the-art approach. By GENIUS-C, we denote the variant of GENIUS augmented with CLIP-SF-based reranking. For text-to-image retrieval, representative methods include GRACE~\cite{li2024generative}, IRGen~\cite{zhang2024irgen}, AVG~\cite{li2024revolutionizing}, and ComGTIR-D~\cite{licomgtir}, with ComGTIR-D serving as the current state-of-the-art.

We evaluate three variants of our proposed model: (1) DrIG: The base generative retriever without additional reranking. (2) DrIG-C and DrIG-LT: Two extensions of DrIG that rerank the top candidates using CLIP-SF and LamRA embeddings, respectively.

\subsection{Main Comparison on M-BEIR (RQ1)}
Table~\ref{tab:mbeir_recall_combined} summarizes the main comparison results on M-BEIR under the two evaluation settings introduced in Section~\ref{sec:dataset}: local-pool retrieval and global-pool retrieval. 
Overall, DrIG consistently improves over the generative baseline GENIUS in both settings, increasing the average score from 29.5 to 38.0 in local-pool retrieval and from 28.6 to 36.4 in global-pool retrieval. 
When combined with dense vector-based reranking, DrIG-C and DrIG-LT further improve the results, showing that generative candidate generation and dense vector-based reranking provide complementary signals. 
In the following subsections, we analyze these results in detail.

\begin{table*}[!t]
\caption{The comparison with state-of-the-art (SOTA) methods under two evaluation settings: local-pool retrieval and global-pool retrieval. Bold values indicate the best performance among all methods, 
and underlined values indicate the best performance within each respective retrieval 
strategy. $\Delta\%$ denotes the relative improvement over the corresponding 
generative baseline.}
\centering
\resizebox{\columnwidth}{!}{%
\footnotesize 
\setlength{\tabcolsep}{3.5pt} 
\renewcommand{\arraystretch}{1.25} 

\begin{tabular}{l l l | c c c | c | c c | c c c}
\toprule
\multicolumn{3}{c|}{\textbf{Retrieval Strategy}} & 
\multicolumn{3}{c|}{\textbf{\shortstack{Dense Vector-based \\ Retrieval}}} & 
\multicolumn{1}{c|}{\textbf{\shortstack{LMM-based \\ Reranking}}} & 
\multicolumn{2}{c|}{\textbf{\shortstack{Generative \\ Retrieval}}} & 
\multicolumn{3}{c}{\textbf{\shortstack{Generative Retrieval + \\ Dense Vector-based Reranking}}} \\

\cmidrule(lr){1-3} \cmidrule(lr){4-6} \cmidrule(lr){7-7} \cmidrule(lr){8-9} \cmidrule(lr){10-12}

\multicolumn{3}{c|}{\textbf{\#Parameters}} & 428M & 447M & 7B & 7B & \textbf{30M} & \textbf{30M} & -- & -- & -- \\
\midrule

\textbf{Task} & \textbf{Dataset} & \textbf{Metric} & CLIP-SF & BLIP-FF & LamRA-ret & LamRA & GENIUS & \textbf{DrIG} {\scriptsize ($\Delta$\%)} & GENIUS-C & \textbf{DrIG-C} {\scriptsize ($\Delta$\%)} & \textbf{DrIG-LT} {\scriptsize ($\Delta$\%)} \\
\midrule

% #####################################################################
% PART 1: Task-Specific Information
% #####################################################################
\multicolumn{12}{c}{\textbf{\textit{(a) Local-pool Retrieval Setting}}} \\
\midrule

% 1. q_t -> c_i
\multirow{3}{*}{\textbf{1}: $q_t \to c_i$} 
& VisualNews & R@5 & \underline{42.4} & 22.8 & 41.6 & \textbf{48.0} & 17.7 & \underline{22.1}{\scriptsize\textcolor{green!50!black}{(+24.9)}} & 26.1 & 33.8{\scriptsize\textcolor{green!50!black}{(+29.5)}} & \underline{33.8}{\scriptsize\textcolor{green!50!black}{(+29.5)}} \\
& MSCOCO & R@5 & 80.6 & 79.4 & \underline{81.5} & \textbf{85.2} & 61.3 & \underline{70.2}{\scriptsize\textcolor{green!50!black}{(+14.5)}} & 66.0 & 78.8{\scriptsize\textcolor{green!50!black}{(+19.4)}} & \underline{79.6}{\scriptsize\textcolor{green!50!black}{(+20.6)}} \\
& Fashion200K & R@10 & 17.9 & 26.1 & \underline{28.5} & \textbf{32.9} & 12.2 & \underline{15.9}{\scriptsize\textcolor{green!50!black}{(+30.3)}} & 15.0 & 19.3{\scriptsize\textcolor{green!50!black}{(+28.7)}} & \underline{22.3}{\scriptsize\textcolor{green!50!black}{(+48.7)}} \\
\midrule

% 2. q_t -> c_t
\textbf{2}: $q_t \to c_t$
& WebQA & R@5 & 84.0 & 79.1 & \underline{86.0} & \textbf{96.7} & 32.6 & \underline{45.3}{\scriptsize\textcolor{green!50!black}{(+39.0)}} & 43.1 & 64.7{\scriptsize\textcolor{green!50!black}{(+50.1)}} & \underline{65.9}{\scriptsize\textcolor{green!50!black}{(+52.9)}} \\
\midrule

% 3. q_t -> (c_i, c_t)
\multirow{2}{*}{\shortstack[l]{\textbf{3}: $q_t$\\ $\to (c_i, c_t)$}}
& EDIS & R@5 & 53.6 & 49.8 & \underline{62.4} & \textbf{75.8} & 35.7 & \underline{38.7}{\scriptsize\textcolor{green!50!black}{(+8.4)}} & 43.3 & 50.4{\scriptsize\textcolor{green!50!black}{(+16.4)}} & \underline{56.1}{\scriptsize\textcolor{green!50!black}{(+29.6)}} \\
& WebQA & R@5 & 78.2 & 78.0 & \underline{81.2} & \textbf{87.7} & 47.3 & \underline{57.6}{\scriptsize\textcolor{green!50!black}{(+21.8)}} & 57.9 & 69.7{\scriptsize\textcolor{green!50!black}{(+20.4)}} & \underline{70.7}{\scriptsize\textcolor{green!50!black}{(+22.1)}} \\
\midrule

% 4. q_i -> c_t
\multirow{3}{*}{\textbf{4}: $q_i \to c_t$} 
& VisualNews & R@5 & \underline{42.5} & 23.0 & 39.6 & \textbf{48.6} & 17.9 & \underline{21.5}{\scriptsize\textcolor{green!50!black}{(+20.1)}} & 25.5 & \underline{33.7}{\scriptsize\textcolor{green!50!black}{(+32.2)}} & 32.7{\scriptsize\textcolor{green!50!black}{(+28.2)}} \\
& MSCOCO & R@5 & \underline{91.7} & 90.7 & 90.6 & \textbf{92.3} & 82.2 & \underline{84.6}{\scriptsize\textcolor{green!50!black}{(+2.9)}} & 89.9 & \underline{92.0}{\scriptsize\textcolor{green!50!black}{(+2.3)}} & 90.0{\scriptsize\textcolor{green!50!black}{(+0.1)}} \\
& Fashion200K & R@10 & 18.1 & 28.5 & \underline{30.4} & \textbf{36.1} & 12.1 & \underline{16.7}{\scriptsize\textcolor{green!50!black}{(+38.0)}} & 16.7 & 20.1{\scriptsize\textcolor{green!50!black}{(+20.4)}} & \underline{24.7}{\scriptsize\textcolor{green!50!black}{(+47.9)}} \\
\midrule

% 5. q_i -> c_i
\textbf{5}: $q_i \to c_i$
& NIGHTS & R@5 & 31.0 & 31.7 & \underline{32.1} & \textbf{33.5} & 9.5 & \underline{17.7}{\scriptsize\textcolor{green!50!black}{(+86.3)}} & 30.0 & 31.7{\scriptsize\textcolor{green!50!black}{(+5.7)}} & \underline{32.0}{\scriptsize\textcolor{green!50!black}{(+6.7)}} \\
\midrule

% 6. (q_i,q_t) -> c_t
\multirow{2}{*}{\shortstack[l]{\textbf{6}: $(q_i,q_t)$\\$ \to c_t$} }
& OVEN & R@5 & 45.8 & 42.6 & \underline{54.1} & \textbf{59.2} & 35.3 & \underline{42.2}{\scriptsize\textcolor{green!50!black}{(+19.5)}} & 38.4 & \underline{51.1}{\scriptsize\textcolor{green!50!black}{(+33.1)}} & 51.0{\scriptsize\textcolor{green!50!black}{(+32.8)}} \\
& InfoSeek & R@5 & 27.2 & 23.2 & \underline{52.1} & \textbf{64.1} & 11.4 & \underline{25.0}{\scriptsize\textcolor{green!50!black}{(+119.3)}} & 19.8 & 35.0{\scriptsize\textcolor{green!50!black}{(+76.8)}} & \underline{44.1}{\scriptsize\textcolor{green!50!black}{(+122.7)}} \\
\midrule

% 7. (q_i,q_t) -> c_i
\multirow{2}{*}{\shortstack[l]{\textbf{7}: $(q_i,q_t) $\\ $\to c_i$}}
& FashionIQ & R@10 & 24.7 & 29.1 & \underline{33.1} & \textbf{37.8} & 12.9 & \underline{19.5}{\scriptsize\textcolor{green!50!black}{(+51.2)}} & 18.3 & 24.0{\scriptsize\textcolor{green!50!black}{(+31.1)}} & \underline{26.5}{\scriptsize\textcolor{green!50!black}{(+44.8)}} \\
& CIRR & R@5 & 44.6 & 50.5 & \underline{53.1} & \textbf{63.3} & 21.8 & \underline{33.7}{\scriptsize\textcolor{green!50!black}{(+54.6)}} & 37.3 & 45.3{\scriptsize\textcolor{green!50!black}{(+21.4)}} & \underline{49.0}{\scriptsize\textcolor{green!50!black}{(+31.4)}} \\
\midrule

% 8. (q_i,q_t) -> (c_i,c_t)
\multirow{2}{*}{\shortstack[l]{\textbf{8}: $(q_i,q_t)$ \\ $\to (c_i,c_t)$}}
& OVEN & R@5 & 68.7 & 56.4 & \underline{76.2} & \textbf{79.2} & 32.6 & \underline{48.7}{\scriptsize\textcolor{green!50!black}{(+49.4)}} & 35.1 & 63.1{\scriptsize\textcolor{green!50!black}{(+79.8)}} & \underline{63.2}{\scriptsize\textcolor{green!50!black}{(+80.1)}} \\
& InfoSeek & R@5 & 48.7 & 30.3 & \underline{63.3} & \textbf{78.3} & 13.2 & \underline{26.8}{\scriptsize\textcolor{green!50!black}{(+103.0)}} & 26.2 & 47.2{\scriptsize\textcolor{green!50!black}{(+80.2)}} & \underline{50.7}{\scriptsize\textcolor{green!50!black}{(+93.5)}} \\
\midrule

% Average (Task-Specific)
-- & Average & -- & 51.3 & 47.9 & \underline{58.1} & \textbf{63.7} & 29.5 & \underline{38.0}{\scriptsize\textcolor{green!50!black}{(+28.8)}} & 37.6 & 48.7{\scriptsize\textcolor{green!50!black}{(+29.5)}} & \underline{50.4}{\scriptsize\textcolor{green!50!black}{(+34.0)}} \\
\midrule

% #####################################################################
% PART 2: Universal Information
% #####################################################################
\multicolumn{12}{c}{\textbf{\textit{(b) Global-pool Retrieval Setting}}} \\
\midrule

% 1. q_t -> c_i
\multirow{3}{*}{\textbf{1}: $q_t \to c_i$}
& VisualNews & R@5 & \underline{42.1} & 22.3 & 41.3 & \textbf{46.9} & 17.7 & \underline{22.0}{\scriptsize\textcolor{green!50!black}{(+24.3)}} & 26.1 & 33.8{\scriptsize\textcolor{green!50!black}{(+29.5)}} & \underline{33.8}{\scriptsize\textcolor{green!50!black}{(+29.5)}} \\
& MSCOCO & R@5 & 71.3 & 65.2 & \underline{75.3} & \textbf{78.0} & 49.9 & \underline{59.2}{\scriptsize\textcolor{green!50!black}{(+18.6)}} & 62.8 & 70.6{\scriptsize\textcolor{green!50!black}{(+12.4)}} & \underline{71.1}{\scriptsize\textcolor{green!50!black}{(+13.2)}} \\
& Fashion200K & R@10 & 17.9 & 26.0 & \underline{28.5} & \textbf{32.5} & 12.2 & \underline{15.8}{\scriptsize\textcolor{green!50!black}{(+29.5)}} & 14.2 & 19.3{\scriptsize\textcolor{green!50!black}{(+35.9)}} & \underline{22.3}{\scriptsize\textcolor{green!50!black}{(+57.0)}} \\
\midrule

% 2. q_t -> c_t
\textbf{2}: $q_t \to c_t$
& WebQA & R@5 & 83.4 & 78.4 & \underline{85.8} & \textbf{96.5} & 31.7 & \underline{44.3}{\scriptsize\textcolor{green!50!black}{(+39.7)}} & 43.0 & 64.3{\scriptsize\textcolor{green!50!black}{(+49.5)}} & \underline{65.1}{\scriptsize\textcolor{green!50!black}{(+51.4)}} \\
\midrule

% 3. q_t -> (c_i, c_t)
\multirow{2}{*}{\shortstack[l]{\textbf{3}: $q_t$\\ $\to (c_i, c_t)$}}
& EDIS & R@5 & 52.7 & 49.2 & \underline{62.3} & \textbf{74.4} & 35.4 & \underline{38.4}{\scriptsize\textcolor{green!50!black}{(+8.5)}} & 43.1 & 49.8{\scriptsize\textcolor{green!50!black}{(+15.5)}} & \underline{56.0}{\scriptsize\textcolor{green!50!black}{(+29.9)}} \\
& WebQA & R@5 & 77.2 & 77.0 & \underline{81.0} & \textbf{87.1} & 47.0 & \underline{56.8}{\scriptsize\textcolor{green!50!black}{(+20.9)}} & 57.5 & 68.5{\scriptsize\textcolor{green!50!black}{(+19.1)}} & \underline{70.0}{\scriptsize\textcolor{green!50!black}{(+21.7)}} \\
\midrule

% 4. q_i -> c_t
\multirow{3}{*}{\textbf{4}: $q_i \to c_t$} 
& VisualNews & R@5 & 38.7 & 21.0 & \underline{39.3} & \textbf{47.6} & 17.7 & \underline{21.1}{\scriptsize\textcolor{green!50!black}{(+19.2)}} & 25.4 & \underline{33.5}{\scriptsize\textcolor{green!50!black}{(+31.9)}} & 32.4{\scriptsize\textcolor{green!50!black}{(+27.6)}} \\
& MSCOCO & R@5 & \underline{91.3} & 89.7 & 90.4 & \textbf{92.4} & 81.5 & \underline{84.5}{\scriptsize\textcolor{green!50!black}{(+3.7)}} & 89.9 & \underline{91.9}{\scriptsize\textcolor{green!50!black}{(+2.2)}} & 90.0{\scriptsize\textcolor{green!50!black}{(+0.1)}} \\
& Fashion200K & R@10 & 18.0 & 27.2 & \underline{30.4} & \textbf{36.6} & 11.7 & \underline{18.0}{\scriptsize\textcolor{green!50!black}{(+53.8)}} & 16.7 & 20.2{\scriptsize\textcolor{green!50!black}{(+21.0)}} & \underline{24.5}{\scriptsize\textcolor{green!50!black}{(+46.7)}} \\
\midrule

% 5. q_i -> c_i
\textbf{5}: $q_i \to c_i$
& NIGHTS & R@5 & 30.9 & 31.6 & \underline{32.1} & \textbf{34.2} & 9.5 & \underline{17.7}{\scriptsize\textcolor{green!50!black}{(+86.3)}} & 30.0 & 31.4{\scriptsize\textcolor{green!50!black}{(+4.7)}} & \underline{31.8}{\scriptsize\textcolor{green!50!black}{(+6.0)}} \\
\midrule

% 6. (q_i,q_t) -> c_t
\multirow{2}{*}{\shortstack[l]{\textbf{6}: $(q_i,q_t) $\\$\to c_t$} }
& OVEN & R@5 & 39.5 & 39.4 & \underline{48.4} & \textbf{54.0} & 34.0 & \underline{40.4}{\scriptsize\textcolor{green!50!black}{(+18.8)}} & 37.8 & 46.3{\scriptsize\textcolor{green!50!black}{(+22.5)}} & \underline{47.1}{\scriptsize\textcolor{green!50!black}{(+24.6)}} \\
& InfoSeek & R@5 & 22.1 & 19.7 & \underline{48.7} & \textbf{58.7} & 9.9 & \underline{22.2}{\scriptsize\textcolor{green!50!black}{(+124.2)}} & 17.7 & 31.3{\scriptsize\textcolor{green!50!black}{(+76.8)}} & \underline{40.3}{\scriptsize\textcolor{green!50!black}{(+127.7)}} \\
\midrule

% 7. (q_i,q_t) -> c_i
\multirow{2}{*}{\shortstack[l]{\textbf{7}: $(q_i,q_t)$\\$ \to c_i$}} 
& FashionIQ & R@10 & 24.4 & 28.8 & \underline{33.1} & \textbf{37.4} & 12.8 & \underline{19.1}{\scriptsize\textcolor{green!50!black}{(+49.2)}} & 18.2 & 23.7{\scriptsize\textcolor{green!50!black}{(+30.2)}} & \underline{26.3}{\scriptsize\textcolor{green!50!black}{(+44.5)}} \\
& CIRR & R@5 & 43.1 & 48.1 & \underline{50.5} & \textbf{59.7} & 21.1 & \underline{31.0}{\scriptsize\textcolor{green!50!black}{(+46.9)}} & 36.6 & 43.5{\scriptsize\textcolor{green!50!black}{(+18.9)}} & \underline{45.4}{\scriptsize\textcolor{green!50!black}{(+24.0)}} \\
\midrule

% 8. (q_i,q_t) -> (c_i,c_t)
\multirow{2}{*}{\shortstack[l]{\textbf{8}: $(q_i,q_t)$ \\ $\to (c_i,c_t)$}}
& OVEN & R@5 & 59.7 & 55.8 & \underline{70.0} & \textbf{72.6} & 36.8 & \underline{47.0}{\scriptsize\textcolor{green!50!black}{(+27.7)}} & 46.4 & 63.0{\scriptsize\textcolor{green!50!black}{(+35.8)}} & \underline{63.7}{\scriptsize\textcolor{green!50!black}{(+37.3)}} \\
& InfoSeek & R@5 & 44.1 & 26.1 & \underline{60.0} & \textbf{74.0} & 12.4 & \underline{25.3}{\scriptsize\textcolor{green!50!black}{(+104.0)}} & 25.3 & 44.4{\scriptsize\textcolor{green!50!black}{(+75.5)}} & \underline{47.8}{\scriptsize\textcolor{green!50!black}{(+88.9)}} \\
\midrule

% Average (Universal)
-- & Average & -- & 48.6 & 45.7 & \underline{56.3} & \textbf{61.4} & 28.6 & \underline{36.4}{\scriptsize\textcolor{green!50!black}{(+27.3)}} & 37.8 & 47.1{\scriptsize\textcolor{green!50!black}{(+24.6)}} & \underline{48.9}{\scriptsize\textcolor{green!50!black}{(+29.4)}} \\
\bottomrule
\end{tabular}
\label{tab:mbeir_recall_combined}
}
\end{table*}

\subsubsection{Local-pool retrieval}
Table~\ref{tab:mbeir_recall_combined}(a) reports the results under the local-pool setting.
Compared with the generative baseline GENIUS, DrIG achieves consistent improvements across all retrieval tasks, increasing the average score from 29.5 to 38.0, corresponding to a relative gain of 28.8\%.
The gains are particularly pronounced on knowledge-intensive and compositional tasks, such as InfoSeek in Task~6, where the score increases from 11.4 to 25.0, and CIRR in Task~7, where the score increases from 21.8 to 33.7. These results suggest that the proposed dual-role identifier is effective for heterogeneous multimodal retrieval. Specifically, the sequential role preserves the autoregressive generation process, while the set-based role introduces a global, prefix-independent relevance signal that helps constrained beam search retain candidates that are semantically compatible with the full query.

Dense vector-based reranking further improves performance, indicating that generative retrieval and dense vector-based retrieval provide complementary signals.
Specifically, DrIG-C and DrIG-LT raise the average score to 48.7 and 50.4, respectively, outperforming both DrIG and GENIUS-C.
This supports a natural two-stage interpretation in which the generative retriever efficiently produces a compact candidate set, and the dense vector-based reranker refines the final order using continuous, fine-grained semantic similarity.
Nevertheless, a clear gap remains between the best DrIG variant and the strongest LMM-based reranker LamRA, especially on text-centric knowledge retrieval.
For example, on WebQA in Task~2 ($q_t \rightarrow c_t$), DrIG-LT achieves 65.9, whereas LamRA reaches 96.7.
This indicates that discretizing dense representations into short identifier sequences may lose information that is important for long, knowledge-dense text matching, and that reranking can mitigate but not fully eliminate this limitation.
\subsubsection{Global-pool retrieval}
Table~\ref{tab:mbeir_recall_combined}(b) reports the results under the global-pool setting. Compared with the local-pool setting, this setting is more challenging because all candidates are searched within a unified pool, requiring the model to infer the target modality, domain, and retrieval intent from the instruction.
Under this more demanding setting, DrIG again consistently outperforms the generative baseline GENIUS across all retrieval tasks, improving the average score from 28.6 to 36.4, with a relative gain of 27.3\%.
Large improvements are observed on knowledge-intensive and multimodal tasks, such as InfoSeek in Task~6, where the score increases from 9.9 to 22.2, and InfoSeek in Task~8, where the score increases from 12.4 to 25.3.
These results indicate that the dual-role identifier remains effective even when the retrieval space contains heterogeneous candidates from different datasets, domains, and modalities. The transition from local-pool to global-pool retrieval causes only a moderate decrease for DrIG, from 38.0 to 36.4 on average.
A similar trend is observed for strong dense vector-based baselines, such as CLIP-SF, which decreases from 51.3 to 48.6, and LamRA-ret, which decreases from 58.1 to 56.3.
This suggests that DrIG is reasonably robust to candidate-pool expansion.
One possible explanation is that the modality-aware identifier and instruction-conditioned decoding process help constrain the search toward target-compatible candidate regions, while the set-based global prior provides an additional relevance signal for distinguishing candidates beyond local prefix likelihoods.

Dense vector-based reranking remains beneficial in the global-pool setting.
DrIG-C and DrIG-LT further improve the average score to 47.1 and 48.9, respectively, clearly outperforming GENIUS-C and surpassing BLIP-FF.
Notably, DrIG-LT also becomes competitive with CLIP-SF, showing that generative candidate generation combined with dense reranking can provide a favorable balance between retrieval effectiveness and scalability.
Nevertheless, DrIG-LT still lags behind LamRA-ret and LamRA, whose average scores are 56.3 and 61.4, respectively.
This gap indicates that although dual-role identifiers improve generative retrieval substantially, discretization still loses fine-grained semantic information, especially for tasks that require precise textual or knowledge-intensive matching.
Overall, the global-pool results confirm the main findings from the local-pool setting: DrIG provides a stronger generative retriever than GENIUS, and hybrid reranking further narrows the effectiveness gap between generative and dense retrieval paradigms.
\subsection{Text-to-image Generative Retrieval (RQ2)}
% text-t-image
\begin{table*}[t]
\caption{The comparison of generative retrieval methods on text-to-image retrieval benchmarks. The best and second-best results for each metric are shown in bold and underlined, respectively.  $\dagger$ indicates zero-shot performance, highlighting the ability of the model to generalize without task-specific fine-tuning.}
\centering
\tiny
\resizebox{0.8\columnwidth}{!}{%
\begin{tabular}{llccc}
\toprule
\textbf{Method} & \textbf{Training Data} & \textbf{Recall@1} & \textbf{Recall@5} & \textbf{Recall@10} \\
\midrule
\multicolumn{5}{c}{\textbf{Flickr30K}} \\
\midrule
GRACE~\cite{li2024generative} & Flickr30K & 37.4 & 59.5 & 66.2 \\
IRGen~\cite{zhang2024irgen} & Flickr30K & 49.0 & 68.9 & 72.5 \\
AVG~\cite{li2024revolutionizing} & Flickr30K & 40.8 & 75.1 & 84.2 \\
ComGTIR-D~\cite{licomgtir} & Flickr30K & 42.6 & 75.7 & 85.3 \\
ComGTIR-DH$_{\mathrm{clip}}$~\cite{licomgtir} & Flickr30K & 68.4 & 86.3 & 90.7 \\
\midrule
\textbf{DrIG}     & M-BEIR    & \textit{59.0}$^{\dagger}$ & \textit{83.1}$^{\dagger}$ & \textit{88.2}$^{\dagger}$ \\
\textbf{DrIG-LT}  & M-BEIR    & \underline{\textit{75.8}}$^{\dagger}$ & \underline{\textit{90.0}}$^{\dagger}$ & \textit{91.6}$^{\dagger}$ \\
\textbf{DrIG}     & Flickr30K & 65.8 & 88.4 & \underline{92.8} \\
\textbf{DrIG-LT}  & Flickr30K & \textbf{76.9} & \textbf{92.5} & \textbf{94.8} \\
\midrule
\multicolumn{5}{c}{\textbf{MSCOCO}} \\
\midrule
GRACE~\cite{li2024generative} & MSCOCO & 16.7 & 39.2 & 50.3 \\
IRGen~\cite{zhang2024irgen} & MSCOCO & 29.6 & 50.7 & 56.3 \\
AVG~\cite{li2024revolutionizing} & MSCOCO & 19.3 & 45.7 & 59.7 \\
ComGTIR-D~\cite{licomgtir} & MSCOCO & 20.6 & 46.7 & 61.0 \\
ComGTIR-DH$_{\mathrm{clip}}$~\cite{licomgtir} & MSCOCO & 45.2 & 64.2 & 72.4 \\
\midrule
\textbf{DrIG}     & M-BEIR & 41.8 & 70.2 & 79.4 \\
\textbf{DrIG-LT}  & M-BEIR & \underline{56.1} & \underline{79.6} & \underline{86.0} \\
\textbf{DrIG}     & MSCOCO & 43.4 & 71.3 & 80.5 \\
\textbf{DrIG-LT}  & MSCOCO & \textbf{56.3} & \textbf{80.1} & \textbf{86.7} \\
\bottomrule
\end{tabular}
\label{tab:gen_retrieval_comparison}
}
\end{table*}

Table~\ref{tab:gen_retrieval_comparison} compares DrIG with representative generative retrieval methods on Flickr30K and MSCOCO, two widely adopted benchmarks for text-to-image retrieval.
These experiments complement the M-BEIR evaluation by examining whether the proposed generative retrieval framework remains effective in the conventional text-to-image setting.
For Flickr30K, we evaluate both cross-benchmark transfer, where DrIG is trained on M-BEIR and directly tested on Flickr30K without Flickr30K-specific fine-tuning, and in-domain training on Flickr30K.
For MSCOCO, since it is already included in M-BEIR, the M-BEIR-trained results should be interpreted as the performance of the universal model on a standard text-to-image benchmark rather than as a strict zero-shot evaluation.

On Flickr30K, the M-BEIR-trained DrIG achieves strong zero-shot performance, reaching 59.0/83.1/ 88.2 in terms of Recall@1/Recall@5/Recall@10, respectively.
This already outperforms several in-domain generative baselines, including GRACE, IRGen, AVG, and ComGTIR-D, indicating that instruction-conditioned training on heterogeneous M-BEIR tasks yields transferable visual-semantic identifiers.
When combined with LamRA-based reranking, DrIG-LT further improves to 75.8/90.0/91.6, surpassing the strongest prior hybrid baseline, ComGTIR-DH$_{\mathrm{clip}}$.
After in-domain training on Flickr30K, DrIG and DrIG-LT further improve to 65.8/88.4/92.8 and 76.9/92.5/94.8, respectively, showing that task-specific fine-tuning can still provide additional benefits on top of the universal retrieval training.

On MSCOCO, DrIG trained on M-BEIR obtains 41.8/70.2/79.4 in terms of Recall@1/Recall@5/ Recall@10, respectively.
Although its Recall@1 is slightly lower than ComGTIR-DH$_{\mathrm{clip}}$, it achieves higher Recall@5 and Recall@10 than all non-reranked and hybrid generative baselines.
With LamRA-based reranking, DrIG-LT reaches 56.1/79.6/86.0, clearly outperforming existing generative retrieval methods across all three metrics.
In-domain training on MSCOCO brings further but relatively modest gains, improving DrIG to 43.4/71.3/80.5 and DrIG-LT to 56.3/80.1/86.7.
This suggests that the universal M-BEIR training already provides a strong foundation for MSCOCO-style text-to-image retrieval, while in-domain fine-tuning mainly refines dataset-specific matching patterns.

The effect of reranking is also consistent across the two benchmarks.
For example, under M-BEIR training, LamRA-based reranking improves DrIG by 16.8 Recall@1 points on Flickr30K and by 14.3 Recall@1 points on MSCOCO.
The gains at Recall@10 are smaller, suggesting that the generative retriever often includes relevant candidates within the generated top-ranked set, while the dense reranker is particularly useful for promoting the most semantically matched image to the top positions.
Overall, these results show that DrIG is not limited to the heterogeneous M-BEIR setting. It also serves as an effective generative candidate generator for standard text-to-image retrieval, and its combination with dense vector-based reranking yields strong performance under both transfer and in-domain evaluation settings.
\subsection{Component Ablation and Representation Analysis (RQ3)}
We next analyze how different components of DrIG contribute to retrieval effectiveness. All ablations are conducted on the base DrIG model without dense vector-based reranking, so that the effect of each component can be isolated within the generative retrieval framework. We evaluate representative tasks from M-BEIR, including text-to-image retrieval, image-to-text retrieval, text-to-text retrieval, text-to-multimodal retrieval, multimodal-to-image retrieval, and image-to-image retrieval.
\subsubsection{Component Ablation}
Table~\ref{tab:ablation_study_full} reports the ablation results.
We consider the following variants: (1) \textit{w/o Set-based Role} removes the set-based interpretation of the identifier, so inference relies only on sequential identifier generation under constrained beam search. (2) \textit{w/o Trie} removes the prefix-tree constraint, allowing the decoder to generate identifiers without prefix-level validity checking. (3) \textit{w/o Trie \& Set-based} removes both the structural validity constraint and the set-based global relevance prior. (4) \textit{w/o Query Augmentation} disables the query--target interpolation strategy used for decoder training. (5) \textit{w/o \(\mathcal{L}_{\mathrm{rank}}\)} removes the discriminative ranking objective in Equation-\ref{rankloss}, leaving only the generative sequence prediction objective. (6) \textit{w/o \(\mathcal{L}^{\mathrm{cl}}_{\mathrm{sr}}\)} removes the contrastive loss in Equation-\ref{eq:contrastive_loss} during identifier construction. (7) Finally, \textit{w/o Modality} removes the dedicated first-level modality codebook and uses only semantic residual codebooks for quantization.

\begin{table*}[!htbp]
\caption{The impact of different components in DrIG. The best and second-best results for each metric are shown in bold and underlined, respectively.}
\centering
\scriptsize
\setlength{\tabcolsep}{1.0pt}
\renewcommand{\arraystretch}{0.95}

\resizebox{\textwidth}{!}{%

\begin{tabular}{l | ccc | ccc | ccc | ccc | ccc | ccc}
\toprule
& \multicolumn{6}{c|}{\textbf{MSCOCO}} 
& \multicolumn{6}{c|}{\textbf{WebQA}} 
& \multicolumn{3}{c|}{\textbf{CIRR}} 
& \multicolumn{3}{c}{\textbf{NIGHTS}} \\

& \multicolumn{3}{c|}{$q_t \to c_i$} & \multicolumn{3}{c|}{$q_i \to c_t$} 
& \multicolumn{3}{c|}{$q_t \to c_t$} & \multicolumn{3}{c|}{$q_t \to (c_i, c_t)$} 
& \multicolumn{3}{c|}{$(q_i, q_t) \to c_i$} 
& \multicolumn{3}{c}{$q_i \to c_i$} \\

\textbf{Component} & R@1 & R@5 & R@10 & R@1 & R@5 & R@10 & R@1 & R@5 & R@10 & R@1 & R@5 & R@10 & R@1 & R@5 & R@10 & R@1 & R@5 & R@10 \\
\midrule
Full          
& \textbf{41.8} & \textbf{70.2} & \textbf{79.4} 
& \textbf{57.8} & \textbf{84.6} & \textbf{91.6} 
& \textbf{20.6} & \textbf{45.2} & \textbf{55.9} 
& \textbf{30.9} & \textbf{57.6} & \textbf{67.3} 
& \textbf{14.4} & \textbf{33.7} & \textbf{45.2} 
& 3.1 & \underline{17.7} & \underline{34.8} \\

w/o Set-based Role   
& \underline{40.8} & \underline{68.6} & 77.9 
& 56.9 & 83.2 & \underline{91.5} 
& 19.7 & \underline{43.7} & \underline{52.8} 
& 29.5 & \underline{55.7} & 65.0 
& 12.5 & 32.2 & 43.3 
& 2.9 & 16.0 & 33.3 \\

w/o Trie 
& 20.6 & 21.6 & 21.6 
& 50.7 & 60.2 & 60.5 
& 18.9 & 29.8 & 30.3 
& 27.9 & 41.2 & 43.1 
& 10.7 & 14.6 & 14.9 
& 2.9 & 14.0 & 19.8 \\

w/o Trie  \& Set-based Role
& 0.1 & 0.1 & 0.1 
& 1.0 & 1.0 & 1.0 
& 0.3 & 0.4 & 0.4 
& 3.6 & 4.6 & 4.6 
& 0.1 & 0.1 & 0.1  
& 0.9 & 1.9 & 2.7 \\

w/o Query augmentation
& 37.1 & 66.8 & 77.7 
& 43.9 & 72.9 & 83.3 
& 14.9 & 36.9 & 49.5 
& 27.4 & 52.2 & 63.7 
& 11.0 & 28.8 & 40.1 
& 2.8 & 15.0 & 29.6 \\

w/o $\mathcal{L}_{\mathrm{rank}}$ in Equation-\ref{rankloss}
& 40.3 & 67.8 & 77.3 
& 56.5 & 83.1 & 90.8 
& \underline{20.1} & 43.4 & 52.0 
& \underline{29.7} & 55.4 & 65.3 
& 12.9 & 31.9 & 42.7 
& \underline{3.2} & 15.8 & 32.9  \\

w/o $\mathcal{L}^{\mathrm{cl}}_{\mathrm{sr}}$ in Equation-\ref{eq:contrastive_loss}
& 0.5 & 1.9 & 3.2 
& 0.5 & 2.5 & 3.9 
& 0.0 & 0.0 & 0.1 
& 0.0 & 0.2 & 0.3 
& 0.0 & 0.1 & 0.4 
& 0.0 & 0.3 & 0.4  \\

w/o Modality 
& 40.6 & \underline{68.6} & \underline{78.1} 
& \underline{57.3} & \underline{84.3} & 91.2 
& 19.7 & 42.3 & 52.3 
& 28.3 & 54.4 & \underline{65.8} 
& \underline{13.7} & \underline{33.4} & \underline{44.9} 
& \textbf{3.7} & \textbf{19.1} & \textbf{35.2} \\

\bottomrule
\end{tabular}%
}
\label{tab:ablation_study_full}
\end{table*}

The most important observation is that contrastive learning before quantization is essential for constructing a usable discrete retrieval space.
Removing \(\mathcal{L}^{\mathrm{cl}}_{\mathrm{sr}}\) causes performance to collapse across almost all tasks.
For example, MSCOCO text-to-image Recall@1 drops from 41.8 to 0.5, MSCOCO image-to-text Recall@1 drops from 57.8 to 0.5, and WebQA text-to-text Recall@1 drops from 20.6 to 0.0.
This indicates that residual quantization alone is insufficient for retrieval-oriented identifier construction.
The pre-quantization embedding space must first be organized according to semantic relevance and retrieval intent. Otherwise, the resulting identifiers cannot reliably preserve query--candidate matching relationships.

The Trie constraint is another critical component.
When the Trie is removed, the decoder is no longer restricted to valid candidate identifiers, leading to large drops in top-\(K\) retrieval performance.
For instance, MSCOCO text-to-image Recall@10 decreases from 79.4 to 21.6, and CIRR Recall@10 decreases from 45.2 to 14.9.
The degradation becomes even more severe when both the Trie constraint and the set-based role are removed, where most metrics approach zero.
These results show that valid-prefix control and global relevance guidance are complementary. The Trie ensures that generated sequences correspond to existing candidates, while the set-based role helps rank valid prefixes using query-level relevance information beyond local token likelihoods.

The set-based role itself provides consistent but more moderate gains.
Removing it reduces MSCOCO text-to-image Recall@1 from 41.8 to 40.8, WebQA text-to-multimodal Recall@5 from 57.6 to 55.7, CIRR Recall@5 from 33.7 to 32.2, and NIGHTS Recall@5 from 17.7 to 16.0.
These improvements are smaller than those brought by contrastive identifier learning or Trie-constrained decoding, but they are systematic across the evaluated tasks.
This supports our motivation that prefix-independent relevance guidance can alleviate the local-optimum risk of purely sequential decoding, especially when multiple valid candidates share similar prefixes.

Query augmentation also contributes substantially to decoder robustness.
Without interpolation-based augmentation, performance drops on all evaluated tasks, with especially large decreases on MSCOCO image-to-text retrieval and WebQA text-to-text retrieval.
For example, MSCOCO image-to-text Recall@1 decreases from 57.8 to 43.9, and WebQA text-to-text Recall@1 decreases from 20.6 to 14.9.
This suggests that query--target interpolation helps the decoder learn a smoother mapping from continuous query embeddings to discrete candidate identifiers, improving generalization to diverse query formulations.

The discriminative ranking loss \(\mathcal{L}_{\mathrm{rank}}\) further improves the alignment between autoregressive generation and retrieval ranking.
Removing it generally lowers top-\(K\) performance, such as MSCOCO text-to-image Recall@5 from 70.2 to 67.8 and WebQA text-to-multimodal Recall@5 from 57.6 to 55.4.
The gains are not uniformly large, and NIGHTS Recall@1 slightly increases from 3.1 to 3.2 without this loss.
This indicates that \(\mathcal{L}_{\mathrm{rank}}\) is mainly helpful for improving ranking consistency across most tasks, while its effect can be task-dependent when the absolute Recall@1 is very low or the candidates are visually fine-grained.

The modality codebook is also generally beneficial, particularly for heterogeneous retrieval tasks.
Removing the first-level modality token reduces performance on most text, image, and multimodal retrieval settings.
For example, WebQA text-to-text Recall@5 decreases from 45.2 to 42.3, and WebQA text-to-multimodal Recall@5 decreases from 57.6 to 54.4.
However, the effect is not universal. On NIGHTS image-to-image retrieval, removing the modality codebook slightly improves Recall@1/Recall@5/Recall@10 from 3.1/17.7/34.8 to 3.7/19.1/35.2.
A possible explanation is that NIGHTS contains only image candidates, so an explicit modality token provides less discriminative benefit and may slightly reduce the capacity available for fine-grained visual similarity.
Overall, the modality codebook is useful for universal multimodal retrieval, but its benefit depends on the degree of modality heterogeneity in the candidate pool.

\begin{figure}[t]
  \centering
  \includegraphics[width=0.6\textwidth]{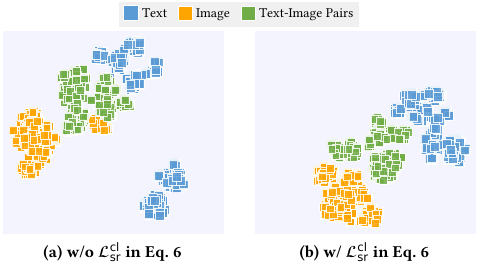}
  \caption{The t-SNE visualizations of multimodal code representations with and without the 
  contrastive loss $\mathcal{L}^{\mathrm{cl}}_{\mathrm{sr}}$. Colors indicate different 
  modalities, including text, image, and image--text pairs.
  }
  \label{fig:tsnechart}
\end{figure}

In summary, the ablation results show that DrIG depends on several complementary components.
Contrastive identifier learning and Trie-constrained decoding are indispensable for building a valid and retrieval-oriented generative index.
The set-based role provides additional prefix-independent relevance guidance, query augmentation improves decoder robustness, the ranking loss better aligns generation with retrieval, and the modality codebook helps organize heterogeneous candidate types.
Together, these components enable DrIG to construct reliable dual-role identifiers for generative universal multimodal retrieval.

\subsubsection{Visualization Analysis}

Figure ~\ref{fig:tsnechart} visualizes the learned code representations with and without the contrastive loss \(L^{\mathrm{cl}}_{\mathrm{sr}}\).
Without this contrastive objective, the representations still exhibit partial modality-related structure, but the clusters are scattered and overlap substantially, especially between text and image--text candidates.
After introducing \(L^{\mathrm{cl}}_{\mathrm{sr}}\), the representations become more compact and more clearly separated across text, image, and image--text modalities.

This visualization provides qualitative evidence for the role of contrastive learning in identifier construction.
By aligning matched query--candidate pairs and separating mismatched candidates before quantization, \(\mathcal{L}^{\mathrm{cl}}_{\mathrm{sr}}\) encourages the residual quantizer to assign identifiers in a more semantically and modality-aware manner.
This observation is consistent with the ablation results in Table~\ref{tab:ablation_study_full}, where removing \(\mathcal{L}^{\mathrm{cl}}_{\mathrm{sr}}\) leads to severe performance degradation.
Therefore, the contrastive objective is not merely an auxiliary training signal. It is a key prerequisite for learning discrete identifiers that preserve retrieval-relevant structure.

\subsection{Effectiveness--Efficiency Trade-off (RQ4)}

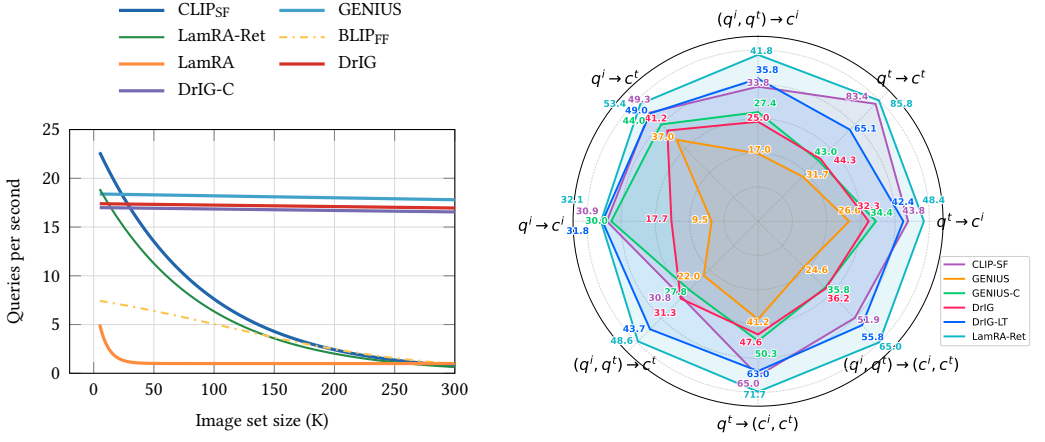
\begin{figure}[t]
\centering

% ============================================================
% Left subfigure: Efficiency curve
% ============================================================
\begin{subfigure}[t]{0.48\textwidth}
\centering
\begin{tikzpicture}
\begin{axis}[
    xlabel={Image set size (K)},
    ylabel={Queries per second},
    xmin=-20, xmax=300,
    ymin=0, ymax=25,
    xtick={0,50,100,150,200,250,300},
    ytick={0,5,10,15,20,25},
    grid=major,
    grid style={line width=.1pt, draw=gray!30},
    legend style={
        at={(0.5,1.08)},
        anchor=south,
        legend columns=2,
        column sep=0.6ex,
        font=\scriptsize,
        draw=none,
        legend cell align={left}
    },
    width=\linewidth,
    height=0.72\linewidth,
    tick label style={
        /pgf/number format/1000 sep=,
        font=\scriptsize
    },
    label style={font=\scriptsize},
    enlargelimits=false,
    enlarge y limits=false,
]

% =========================
% Colors
% =========================
\definecolor{coldblue}{RGB}{33,102,172}
\definecolor{cyanblue}{RGB}{67,162,202}
\definecolor{coldgreen}{RGB}{35,139,69}
\definecolor{warmyellow}{RGB}{254,196,79}
\definecolor{warmorange}{RGB}{253,141,60}
\definecolor{hotred}{RGB}{215,48,39}
\definecolor{purpleviolet}{RGB}{117,107,177}

% -------- 1) CLIP_SF --------
\addplot[
    domain=5:300,
    samples=300,
    color=coldblue,
    very thick,
    smooth
] {24 * exp(-0.01148 * x)};
\addlegendentry{CLIP$_{\mathrm{SF}}$}

% -------- 2) GENIUS --------
\addplot[
    domain=5:300,
    samples=300,
    color=cyanblue,
    very thick,
    smooth
] {18.4 - 0.002*x};
\addlegendentry{GENIUS}

% -------- 3) LamRA-Ret --------
\addplot[
    domain=5:300,
    samples=300,
    color=coldgreen,
    thick,
    smooth
] {20 * exp(-0.01148 * x)};
\addlegendentry{LamRA-Ret}

% -------- 4) BLIP_FF --------
\addplot[
    domain=5:300,
    samples=400,
    color=warmyellow,
    thick,
    dashdotted,
    smooth
] { 1.5 / (0.04545 * exp(0.01148 * x) + 0.1538) };
\addlegendentry{BLIP$_{\mathrm{FF}}$}

% -------- 5) LamRA --------
\pgfmathsetmacro{\eps}{2.1}
\pgfmathsetmacro{\k}{ln(4/\eps)/5}

\addplot[
  domain=5:300,
  samples=400,
  color=warmorange,
  very thick,
  smooth
] { 1 + 4*exp(-\k*(x-5)) };
\addlegendentry{LamRA}

% -------- 6) DrIG --------
\addplot[
    domain=5:300,
    samples=300,
    color=hotred,
    very thick,
    smooth
] {17.4 - 0.0015*x};
\addlegendentry{DrIG}

% -------- 7) DrIG-C --------
\addplot[
    domain=5:300,
    samples=300,
    color=purpleviolet,
    very thick,
    smooth
] {17.0 - 0.0015*x};
\addlegendentry{DrIG-C}

\end{axis}
\end{tikzpicture}
\caption{The efficiency under varying candidate-pool sizes.}
\label{fig:efficiency_curve}
\end{subfigure}
\hfill
% ============================================================
% Right subfigure: Radar plot
% ============================================================
\begin{subfigure}[t]{0.5\textwidth}
\centering

\includegraphics[
    width=\linewidth,
    trim=0 0 0 0,
    clip
]{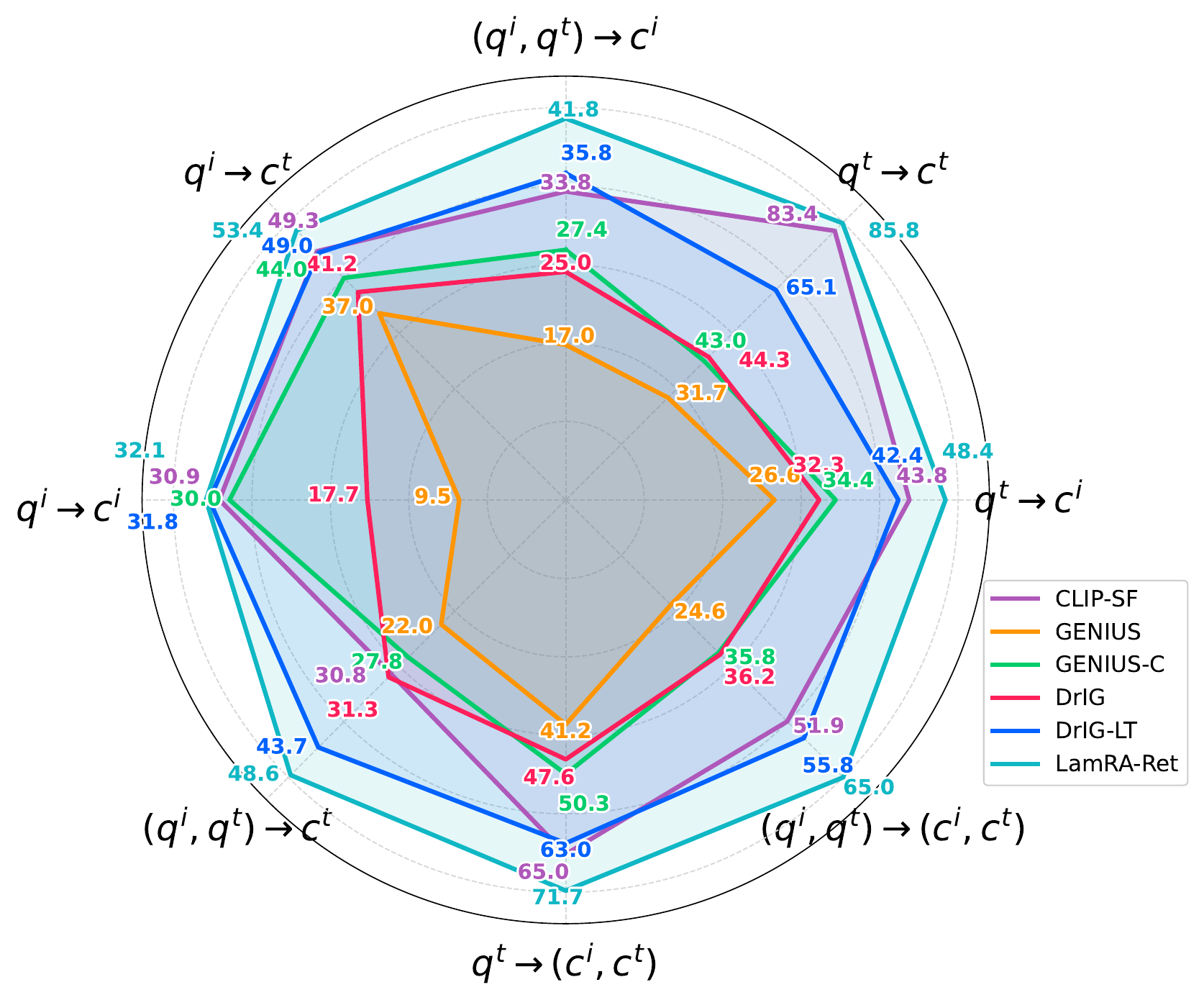}

\caption{The retrieval effectiveness across M-BEIR task types.}
\label{fig:efficiency_radar}
\end{subfigure}

\caption{The effectiveness--efficiency comparison across representative retrieval methods.}
\label{fig:efficiency_and_radar}
\end{figure}

Figure ~\ref{fig:efficiency_and_radar} compares representative retrieval methods from both efficiency and effectiveness perspectives.
Figure ~\ref{fig:efficiency_curve} reports the online throughput, measured in queries per second (QPS), as the image candidate pool increases from 5K to 300K on a single A100 GPU.
Under this evaluation setting, generative retrieval methods exhibit substantially more stable throughput than dense retrieval baselines.
GENIUS and DrIG maintain nearly flat QPS curves as the candidate pool grows, because their main online cost comes from autoregressive decoding over a fixed-length identifier space rather than scoring every candidate in the pool.
DrIG-C, which adds dense vector-based reranking over a fixed-size generated candidate set, also preserves this favorable scaling trend.
In contrast, dense retrieval baselines such as CLIP-SF, BLIP-FF, and LamRA-ret show decreasing throughput as the candidate pool becomes larger, since their retrieval process requires similarity computation over an expanding set of candidate embeddings in this implementation.
LamRA has the lowest throughput, reflecting the additional cost of LMM-based reranking.

Efficiency alone, however, is insufficient for evaluating retrieval systems.
Figure ~\ref{fig:efficiency_radar} therefore compares retrieval effectiveness across different task types.
DrIG consistently improves over GENIUS while retaining the scalability advantage of generative retrieval.
For example, in the global-pool setting, DrIG improves the 
$q^t\rightarrow c^t$ task from 31.7 to 44.3 and the averaged 
multimodal-to-multimodal task $(q^i,q^t)\rightarrow(c^i,c^t)$ from 24.6 to 36.2.
These gains indicate that the proposed dual-role identifiers improve the quality of generated candidates without sacrificing the efficient decoding behavior of generative retrieval. Hybrid reranking provides a further operating point in the effectiveness--efficiency trade-off.
Compared with pure DrIG, DrIG-LT substantially improves effectiveness by using dense vector-based reranking to refine the generated candidate list, raising the average global-pool score from 36.4 to 48.9.
This confirms that generative retrieval and dense vector-based reranking play complementary roles. The generator efficiently narrows the search space to a compact set of candidates, while the reranker performs fine-grained semantic matching within this set. Nevertheless, the strongest dense and LMM-based methods, such as LamRA-ret and LamRA, still achieve higher absolute effectiveness, indicating that discretization can lose information that is useful for fine-grained ranking.
Overall, DrIG offers a favorable trade-off. It is substantially more effective than prior generative retrieval methods while preserving their scalability, and its hybrid variants allow users to exchange additional reranking cost for improved accuracy.

\subsection{Design and Hyperparameter Sensitivity Analysis}
Beyond the main research questions, we further analyze how key design choices and hyperparameters affect DrIG.
Unless otherwise specified, the analyses are conducted with the base DrIG model under the local-pool setting. We focus on five factors: residual-quantization codebook configuration, beam size, reranking depth, decoder backbone, and the weight of the set-based global prior.
\subsubsection{Codebook Configuration}
\begin{table*}[t]
\caption{The ablation study on codebook configurations (Levels $L \times$ Vocab $K$). QPS denotes the average number of queries processed per second across all tasks. Bold values indicate the best performance for each metric. }
\centering
\resizebox{\columnwidth}{!}{%
\small 
\setlength{\tabcolsep}{2.0pt} 

\begin{tabular}{c | ccc | ccc | ccc | ccc | ccc | ccc | c}
\toprule
& \multicolumn{6}{c|}{\textbf{MSCOCO}} & \multicolumn{6}{c|}{\textbf{WebQA}} & \multicolumn{3}{c|}{\textbf{CIRR}} & \multicolumn{3}{c|}{\textbf{NIGHTS}} & \textbf{QPS} \\
& \multicolumn{3}{c|}{$q_t \to c_i$} 
& \multicolumn{3}{c|}{$q_i \to c_t$} 
& \multicolumn{3}{c|}{$q_t \to c_t$} 
& \multicolumn{3}{c|}{$q_t \to (c_i, c_t)$} 
& \multicolumn{3}{c|}{$(q_i, q_t) \to c_i$} & \multicolumn{3}{c}{$q_i \to c_i$} & Ave. \\
\textbf{$L \times K$} & R@1 & R@5 & R@10 & R@1 & R@5 & R@10 & R@1 & R@5 & R@10 & R@1 & R@5 & R@10 & R@1 & R@5 & R@10 & R@1 & R@5 & R@10 & (/s) \\
\midrule
% K = 4096 组
4 $\times$ 4096 & 32.3 & 61.3 & 73.1 & 49.4 & 79.0 & 88.2 & 7.3 & 20.9 & 31.0 & 14.3 & 36.1 & 48.4 & 9.0 & 24.7 & 34.8 & 3.1 & 14.8 & 26.0 & 63.3 \\
6 $\times$ 4096 & 38.4 & 66.6 & 77.0 & 54.8 & 82.8 & 90.5 & 15.7 & 35.9 & 46.0 & 24.9 & 53.7 & 64.0 & 11.7 & 31.1 & 42.3 & 3.1 & 16.7 & 32.1 & 47.0 \\
\rowcolor{gray!20}
8 $\times$ 4096 & \textbf{41.8} & \textbf{70.2} & \textbf{79.4} & \textbf{57.8} & \textbf{84.6} & \textbf{91.6} & \textbf{20.6} & \textbf{45.2} & \textbf{55.9} & \textbf{30.9} & \textbf{57.6} & 67.3 & \textbf{14.4} & 33.7 & \textbf{45.2} & 3.1 & 17.7 & \textbf{34.8} & 37.6 \\
\midrule
% K = 2048 组
4 $\times$ 2048 & 30.5 & 59.3 & 71.1 & 45.9 & 76.5 & 86.5 & 6.7 & 20.0 & 30.0 & 12.9 & 32.9 & 44.8 & 8.1 & 24.1 & 34.3 & 3.6 & 15.0 & 27.7 & \textbf{63.8} \\
6 $\times$ 2048 & 36.5 & 65.8 & 76.4 & 53.2 & 82.3 & 90.3 & 14.1 & 33.8 & 44.8 & 23.6 & 49.3 & 60.4 & 11.9 & 30.0 & 41.2 & \textbf{3.9} & 16.4 & 31.4 & 47.4 \\
8 $\times$ 2048 & 41.0 & 69.4 & 79.0 & 56.5 & 84.1 & 91.3 & 20.0 & 43.3 & 52.8 & 29.9 & 55.8 & \textbf{67.8} & 13.7 & \textbf{33.8} & \textbf{45.2} & \textbf{3.9} & \textbf{18.1} & 33.4 & 37.8 \\
\midrule
% K = 1024 组
4 $\times$ 1024 & 28.3 & 56.9 & 68.8 & 42.1 & 74.2 & 85.3 & 6.0 & 17.4 & 25.9 & 10.7 & 29.0 & 39.2 & 7.7 & 21.7 & 31.3 & 2.6 & 14.6 & 27.3 & 36.7 \\
6 $\times$ 1024 & 35.6 & 64.8 & 75.5 & 51.8 & 80.5 & 89.2 & 15.1 & 32.8 & 42.9 & 22.0 & 47.5 & 58.8 & 10.8 & 28.8 & 40.1 & 3.1 & 15.7 & 31.9 & 27.6 \\
8 $\times$ 1024 & 39.8 & 68.4 & 78.3 & 55.5 & 83.4 & 90.7 & 19.3 & 42.6 & 51.2 & 27.7 & 55.5 & 66.6 & 12.5 & 31.8 & 43.8 & 3.0 & 17.4 & 34.0 & 37.9 \\
\midrule
% K = 256 组
4 $\times$ 256  & 22.6 & 49.3 & 62.4 & 35.1 & 65.1 & 78.1 & 3.0 & 10.7 & 14.9 & 5.1 & 14.1 & 21.1 & 5.0 & 15.6 & 23.8 & 2.1 & 9.7 & 18.7 & 39.1 \\
6 $\times$ 256  & 31.7 & 60.4 & 71.9 & 47.4 & 77.4 & 87.0 & 8.6 & 25.8 & 35.2 & 15.4 & 36.9 & 47.9 & 7.9 & 24.3 & 34.8 & 3.0 & 16.4 & 29.9 & 48.5 \\
8 $\times$ 256  & 36.1 & 65.1 & 75.4 & 52.7 & 80.7 & 89.3 & 15.4 & 36.1 & 45.6 & 22.7 & 47.5 & 59.4 & 11.0 & 28.8 & 39.4 & 3.4 & 17.6 & 31.5 & 26.3 \\
\bottomrule
\end{tabular}
\label{tab:rq_config_full_ablation}
}
\end{table*}

Table~\ref{tab:rq_config_full_ablation} reports the effect of different residual-quantization configurations by varying the number of quantization levels \(L\) and the semantic codebook vocabulary size \(K\).
The first-level modality codebook is kept fixed, while the remaining codebooks are used to encode semantic residuals.
Overall, increasing the quantization depth brings the most consistent improvement.
For example, when \(K=4096\), increasing \(L\) from 4 to 8 improves MSCOCO text-to-image Recall@1 from 32.3 to 41.8 and WebQA text-to-text Recall@1 from 7.3 to 20.6.
This indicates that longer identifiers provide more capacity for capturing fine-grained multimodal semantics. Increasing the vocabulary size also tends to improve retrieval effectiveness, especially when the identifier is sufficiently deep.
For instance, under \(L=8\), the configuration \(8\times4096\) achieves the best performance on most MSCOCO and WebQA metrics.
However, the benefit of a larger vocabulary is not uniform across all tasks.
On visually fine-grained datasets such as CIRR and NIGHTS, \(8\times2048\) performs slightly better on several metrics, including CIRR Recall@5 and NIGHTS Recall@1/Recall@5.
This suggests that an excessively large codebook is not always optimal. For tasks with many visually similar candidates, a moderately sized vocabulary may provide useful regularization.

The efficiency trend is also task- and implementation-dependent.
Increasing the number of quantization levels generally reduces throughput because the decoder must generate longer identifiers.
For example, with \(K=4096\), QPS decreases from 63.3 for \(4\times4096\) to 37.6 for \(8\times4096\).
In contrast, the effect of vocabulary size on QPS is less monotonic, since Trie-constrained decoding only expands valid prefixes rather than all tokens in the vocabulary.
Therefore, codebook design introduces a trade-off between identifier expressiveness and inference efficiency.
We use \(8\times4096\) as the default setting because it provides the strongest and most stable overall effectiveness while maintaining acceptable retrieval throughput.

\subsubsection{Beam Size}
% -------------------------
% Table 9: Beam size ablation
% -------------------------
\begin{table*}[t]
\caption{The comprehensive beam size ablation on decoding settings. QPS represents the average queries per second across all tasks. Bold values indicate the best performance for each metric. }
\centering
\resizebox{\columnwidth}{!}{%
\small % 使用小号字体
\setlength{\tabcolsep}{2.0pt} % 微调列间距以适应 20 列数据

\begin{tabular}{c | ccc | ccc | ccc | ccc | ccc | ccc | c}
\toprule
& \multicolumn{6}{c|}{\textbf{MSCOCO}} & \multicolumn{6}{c|}{\textbf{WebQA}} & \multicolumn{3}{c|}{\textbf{CIRR}} & \multicolumn{3}{c|}{\textbf{NIGHTS}} & \textbf{QPS} \\
& \multicolumn{3}{c|}{$q_t \to c_i$} 
& \multicolumn{3}{c|}{$q_i \to c_t$} 
& \multicolumn{3}{c|}{$q_t \to c_t$} 
& \multicolumn{3}{c|}{$q_t \to (c_i, c_t)$} 
& \multicolumn{3}{c|}{$(q_i, q_t) \to c_i$} & \multicolumn{3}{c}{$q_i \to c_i$} & Ave. \\
\textbf{Beam} & R@1 & R@5 & R@10 & R@1 & R@5 & R@10 & R@1 & R@5 & R@10 & R@1 & R@5 & R@10 & R@1 & R@5 & R@10 & R@1 & R@5 & R@10 & (/s) \\
\midrule
1  & 25.1 & 25.1 & 25.1 & 43.4 & 43.5 & 43.5 & 8.1  & 8.2  & 8.2  & 14.9 & 15.9 & 15.9 & 8.1  & 8.2  & 8.2  & 3.1 & 3.3 & 3.3 & 972.6 \\
5  & 38.9 & 57.6 & 57.6 & 57.2 & 79.8 & 80.1 & 17.2 & 30.9 & 31.0 & 26.5 & 43.4 & 44.6 & 13.5 & 26.9 & 27.2 & \textbf{3.4} & 17.3 & 18.9 & 275.2 \\
10 & 40.8 & 65.0 & 70.3 & 57.6 & 83.1 & 89.2 & 19.7 & 38.1 & 43.0 & 29.4 & 51.0 & 58.0 & 13.8 & 31.2 & 39.3 & 3.2 & 17.8 & 32.8 & 162.8 \\
20 & 41.5 & 68.4 & 76.7 & 57.8 & 84.1 & 91.0 & 20.4 & 42.8 & 50.6 & 30.5 & 55.2 & 63.5 & 14.4 & 32.8 & 43.2 & 3.2 & 18.0 & 34.5 & 87.9 \\
30 & 41.7 & 69.5 & 78.4 & \textbf{57.9} & 84.6 & 91.4 & 20.6 & 44.3 & 53.5 & 30.8 & 56.7 & 66.0 & 14.5 & 33.2 & 44.6 & 3.1 & 17.7 & 34.4 & 61.3 \\
40 & \textbf{41.8} & 69.9 & 79.1 & 57.8 & \textbf{84.7} & \textbf{91.6} & \textbf{20.6} & \textbf{45.3} & 55.2 & \textbf{30.9} & 57.1 & 66.6 & \textbf{14.6} & 33.4 & 44.7 & 3.1 & \textbf{17.7} & \textbf{34.9} & 46.6 \\
\rowcolor{gray!20}
50 & \textbf{41.8} & \textbf{70.2} & \textbf{79.4} & 57.8 & 84.6 & \textbf{91.6} & \textbf{20.6} & 45.2 & \textbf{55.9} & 30.9 & \textbf{57.6} & \textbf{67.3} & 14.4 & \textbf{33.7} & \textbf{45.2} & 3.1 & 17.7 & 34.8 & \textbf{37.6} \\
\bottomrule
\end{tabular}
\label{tab:beam_size_ablation_full}
}
\end{table*}

Table~\ref{tab:beam_size_ablation_full} studies the effect of beam size during Trie-constrained decoding.
Increasing the beam size mainly improves candidate coverage by allowing the decoder to explore more valid identifier paths.
The improvement is especially large when moving from greedy decoding to moderate beam search.
For example, MSCOCO text-to-image Recall@10 increases from 25.1 with beam size 1 to 70.3 with beam size 10, showing that greedy decoding is highly vulnerable to early prefix-level decisions.

The performance gain becomes smaller as the beam size continues to increase.
Most metrics improve rapidly from beam size 1 to 20 and then gradually saturate between 30 and 50.
This pattern suggests that a moderate beam is already sufficient to preserve most relevant candidate paths, while a very large beam mainly refines the top-ranked candidate set.
The improvement is not strictly monotonic for every metric.
For instance, NIGHTS obtains its best Recall@1 at beam size 5, and several Recall@1 values fluctuate slightly at larger beam sizes.
This is expected because a larger beam improves coverage but does not necessarily improve the final top-1 ranking, especially on visually fine-grained tasks with many similar candidates.

Beam size also has a substantial impact on inference efficiency.
QPS decreases from 972.6 with beam size 1 to 37.6 with beam size 50.
Thus, beam size controls a direct effectiveness--efficiency trade-off.
We set the default beam size to 50 in the main experiments to maximize retrieval effectiveness, while beam sizes 20 or 30 provide a practical alternative when lower latency is required.

\subsubsection{Reranking Depth}
%==test recall plot chart
\begin{figure*}[t]
  \centering
  
  \includegraphics[width=1.0\textwidth]{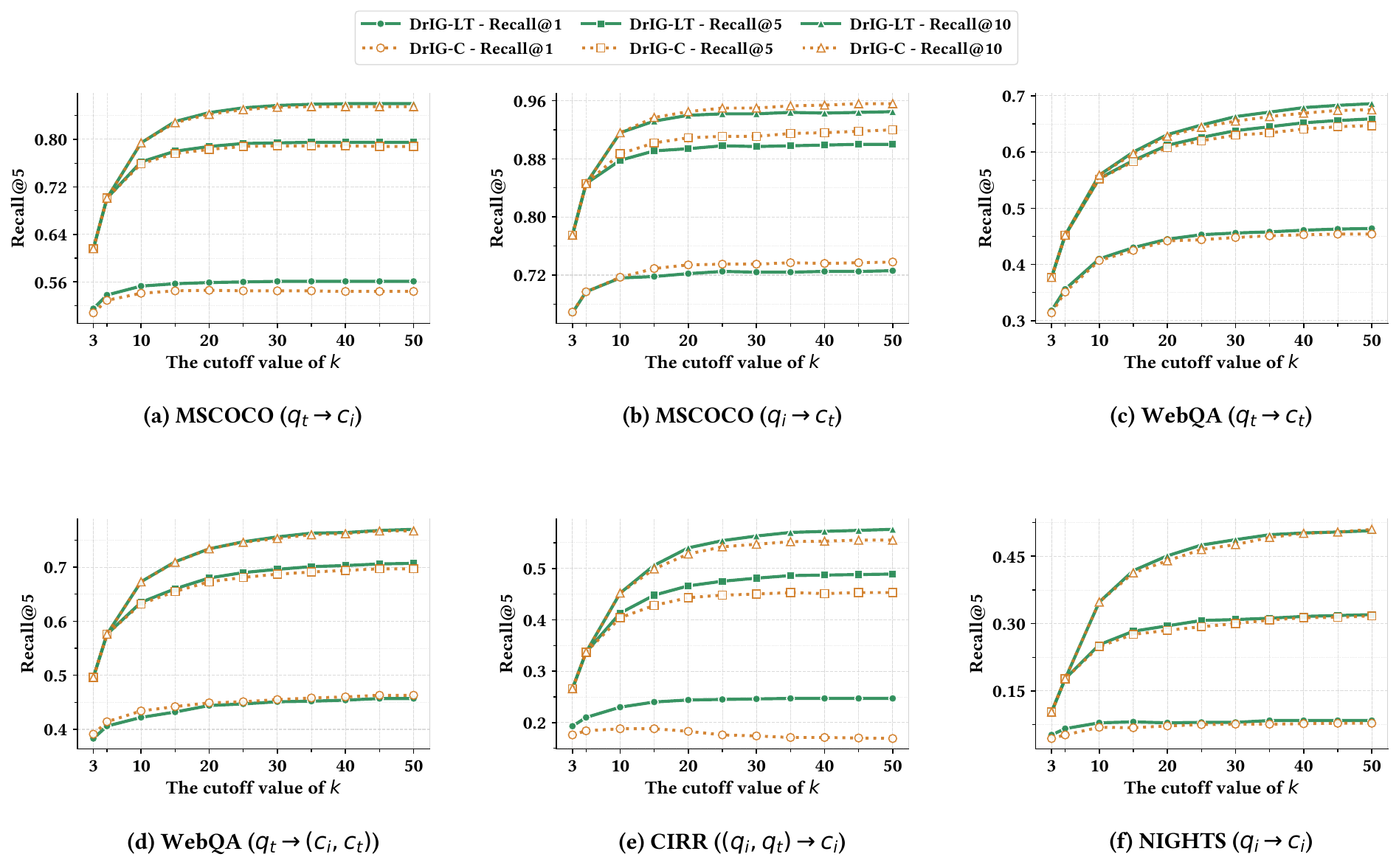}
\caption{The comprehensive ablation study of reranking depth $k$ across different datasets. Recall is reported for $k \in \{3, 5, \dots, 50\}$, highlighting the impact of retrieval-rerank trade-offs in representative tasks.}
  \label{fig:rerank_ablation}
\end{figure*}
%== end of test recall plot chart

Figure ~\ref{fig:rerank_ablation} analyzes the effect of reranking depth \(k\) for the hybrid variants DrIG-C and DrIG-LT.
Here, \(k\) denotes the number of generated candidates passed to the dense vector-based reranker.
The results show that reranking is most beneficial when \(k\) increases from a very small value to a moderate range.
This is because the reranker can only refine the order of candidates that have already been generated by DrIG.
When \(k\) is too small, the generated candidate set may not contain enough relevant or near-relevant candidates for reranking to be effective.
This pattern is especially clear on visually challenging tasks.
For example, on NIGHTS, DrIG-LT obtains a large improvement in Recall@5 when \(k\) increases from 3 to 10, indicating that a small reranking pool is insufficient for image-to-image retrieval, where many candidates are visually similar.
After the main relevant candidates are included, further increasing \(k\) leads to diminishing returns.
Similar saturation trends can be observed on MSCOCO, WebQA, and CIRR.
Across most tasks, DrIG-LT outperforms DrIG-C, showing that LamRA-based embeddings provide stronger fine-grained reranking signals than CLIP-SF embeddings.

These results confirm the complementary roles of generative retrieval and dense vector-based reranking.
The generative retriever efficiently narrows the search space, while the reranker improves the final ordering within the generated candidate set.
In the main experiments, we use \(k=50\) to maximize effectiveness.
For efficiency-sensitive applications, however, moderate values such as \(k=20\) or \(k=30\) can provide a better trade-off between accuracy and reranking cost.
\subsubsection{Decoder Backbone}
% -------------------------
% Table 11: encoder ablation
% -------------------------

\begin{table*}[t]
\caption{The ablation results of model backbone combinations. Bold values indicate the best performance for each metric. }
\centering
\resizebox{\columnwidth}{!}{%
\small % 使用小号字体
\setlength{\tabcolsep}{1.5pt} 

\begin{tabular}{l | c | ccc | ccc | ccc | ccc | ccc | ccc}
\toprule
& & \multicolumn{6}{c|}{\textbf{MSCOCO}} 
& \multicolumn{6}{c|}{\textbf{WebQA}} 
& \multicolumn{3}{c|}{\textbf{CIRR}} 
& \multicolumn{3}{c}{\textbf{NIGHTS}} \\

& 
& \multicolumn{3}{c|}{$q_t \to c_i$} 
& \multicolumn{3}{c|}{$q_i \to c_t$} 
& \multicolumn{3}{c|}{$q_t \to c_t$} 
& \multicolumn{3}{c|}{$q_t \to (c_i, c_t)$} 
& \multicolumn{3}{c|}{$(q_i, q_t) \to c_i$} 
& \multicolumn{3}{c}{$q_i \to c_i$} 
\\

\textbf{Model} & \textbf{\# Params} & R@1 & R@5 & R@10 & R@1 & R@5 & R@10 & R@1 & R@5 & R@10 & R@1 & R@5 & R@10 & R@1 & R@5 & R@10 & R@1 & R@5 & R@10 \\
\midrule

\rowcolor{gray!20}
T5-small & 30M & 41.8 & 70.2 & 79.5 & 57.8 & 84.6 & 91.6 & 20.7 & 45.3 & 56.0 & 30.9 & 57.6 & 67.3 & \textbf{14.4} & \textbf{33.7} & \textbf{45.2} & \textbf{3.1} & \textbf{17.7} & \textbf{34.8} \\

T5-base  & 110M & \textbf{42.5} & \textbf{70.6} & \textbf{80.0} & 58.4 & \textbf{85.3} & 91.8 & 23.2 & 47.1 & \textbf{57.6} & \textbf{32.1} & 58.2 & \textbf{68.6} & 11.4 & 30.5 & 42.2 & 2.3 & 14.1 & 31.1 \\

T5-large & 400M & 42.1 & 70.4 & 79.9 & \textbf{58.5} & 84.8 & \textbf{92.0} & \textbf{24.1} & \textbf{47.7} & 57.3 & \textbf{32.1} & \textbf{58.6} & 68.4 & 10.4 & 27.2 & 39.2 & 1.7 & 11.4 & 27.6 \\
\bottomrule
\end{tabular}
\label{tab:ablation_backbones_final}
}
\end{table*}

Table~\ref{tab:ablation_backbones_final} compares different T5 decoder backbones while keeping the multimodal encoder and identifier construction procedure unchanged.
Increasing the decoder size brings moderate gains on several text-centric or knowledge-intensive tasks.
For example, T5-large improves WebQA text-to-text Recall@1 from 20.7 to 24.1, and T5-base slightly improves MSCOCO text-to-image Recall@1 from 41.8 to 42.5.
These results suggest that a larger decoder can better model semantic and textual patterns during identifier generation. 
However, larger decoders do not improve all tasks.
On visually grounded and compositional retrieval tasks, T5-small is more stable.
For example, on CIRR, T5-small achieves 14.4/33.7/45.2 in Recall@1/Recall@5/Recall@10, whereas T5-large drops to 10.4/27.2/39.2.
A similar trend appears on NIGHTS image-to-image retrieval.
This indicates that simply increasing decoder capacity is not always beneficial for generative multimodal retrieval.
For tasks that depend heavily on fine-grained visual similarity, the bottleneck may lie more in identifier construction and visual-semantic discretization than in the language modeling capacity of the decoder.

Considering both retrieval effectiveness and model efficiency, we use T5-small as the default decoder backbone.
It has only 30M parameters and provides the most balanced cross-task performance, making it suitable for scalable generative retrieval.
\subsubsection{Global Prior Weight $\lambda$}
\begin{figure}[t]
  \centering
  \includegraphics[width=0.9\textwidth]{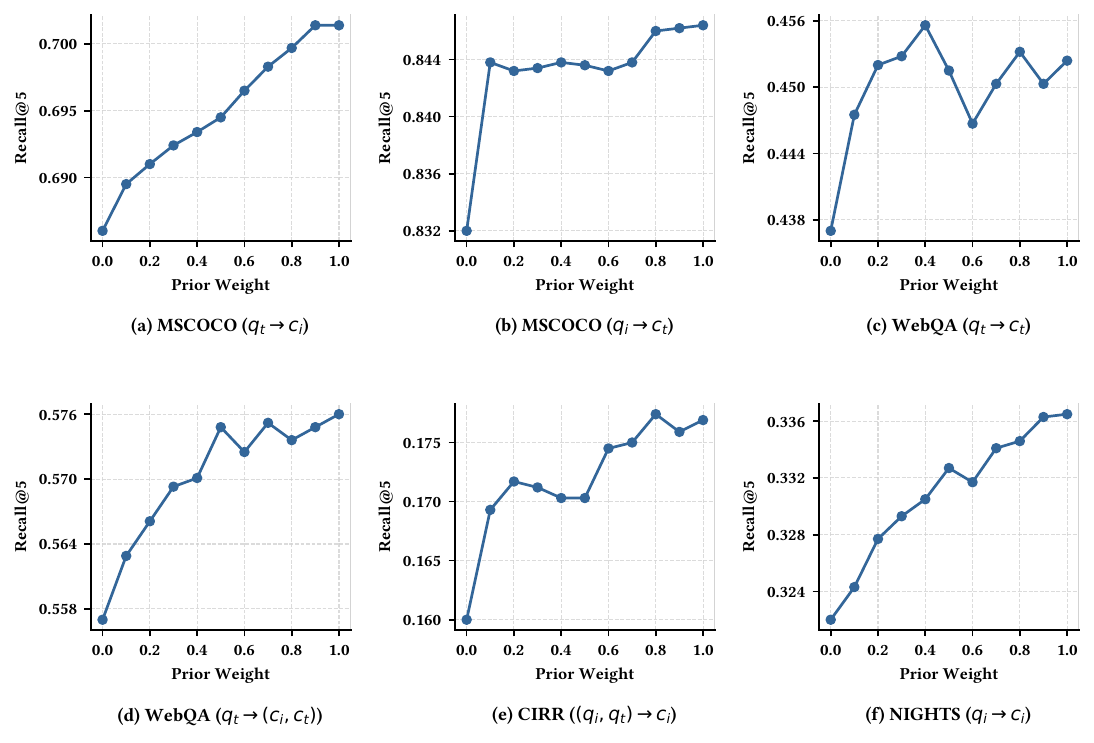}

\caption{The effect of the global prior weight $\lambda$ on Recall@5.}
  \label{fig:lambda_ablation}
\end{figure}

Figure ~\ref{fig:lambda_ablation} studies the effect of the global prior weight \(\lambda\) in Equation-\ref{Eq_g_obj}.
This parameter controls the contribution of the set-based global relevance prior during Trie-constrained beam search.
When \(\lambda=0\), inference relies only on the sequential decoding score and the Trie constraint.
When \(\lambda>0\), the decoder additionally uses the prefix-independent relevance signal derived from the set-based role of the identifier.

The results show that introducing the global prior consistently improves Recall@5 over \(\lambda=0\) on most tasks.
For MSCOCO text-to-image, WebQA text-to-multimodal retrieval, NIGHTS image-to-image retrieval, and CIRR composed image retrieval, performance generally increases as \(\lambda\) becomes larger.
This confirms that the set-based role provides useful global guidance beyond local autoregressive likelihoods.
By scoring candidates reachable from a prefix using the full query representation, the global prior helps the decoder retain semantically promising branches that may otherwise be pruned early.
The optimal value of \(\lambda\) is task-dependent.
For example, text-centric retrieval can exhibit small fluctuations when the prior weight becomes large, suggesting that the set-based prior and sequential decoder score may have different calibration behaviors across modalities and tasks.
Nevertheless, \(\lambda=1.0\) provides strong and stable performance across most evaluated settings.
We therefore use \(\lambda=1.0\) as the default value in the main experiments, while noting that task-specific tuning may yield further improvements in specialized retrieval scenarios.

\subsection{Qualitative Case Studies}

%== case
%== case
%== case
\begin{figure}[t]
  \centering
  \makebox[\textwidth][c]{%
    \includegraphics[
      width=\textwidth,
      keepaspectratio
    ]{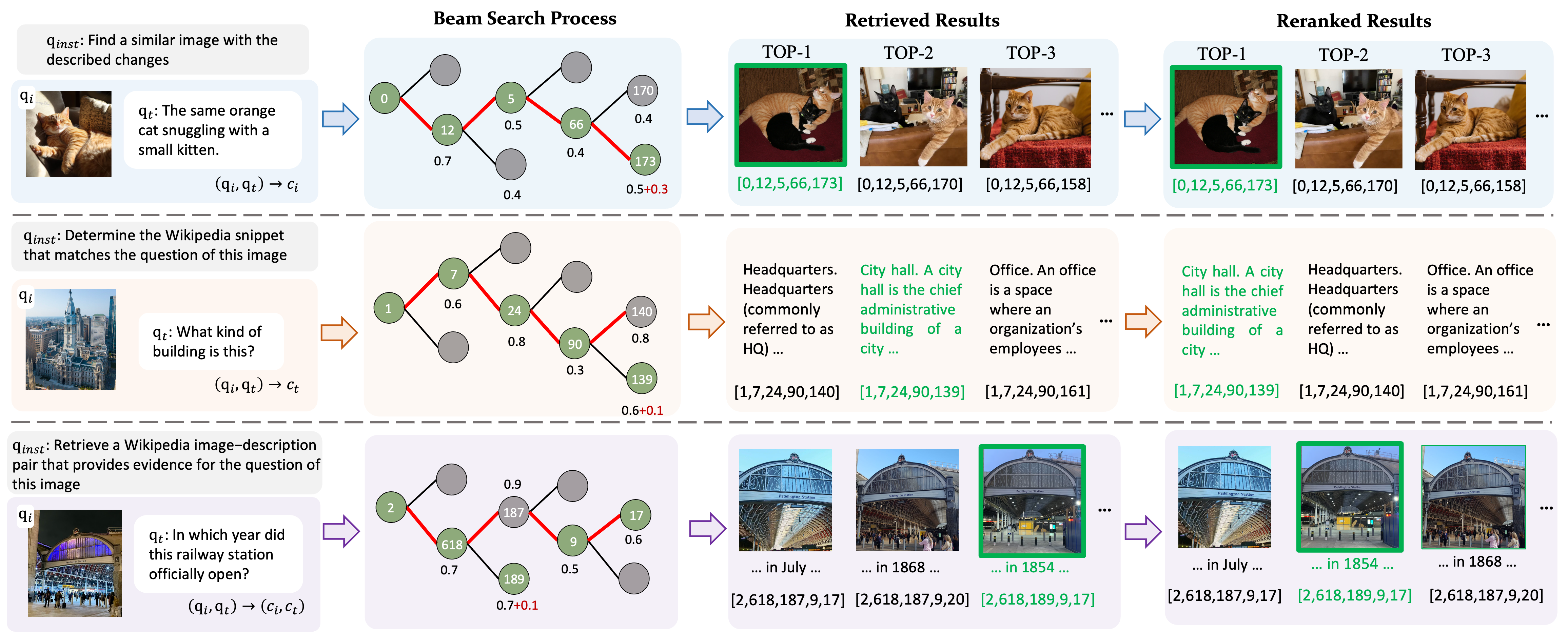}
  }
  \caption{The qualitative case studies of DrIG.}
  \label{fig:case_study}
\end{figure}

Figure ~\ref{fig:case_study} presents three representative cases that illustrate how DrIG behaves during dual-guided decoding and hybrid reranking.
The red path denotes the identifier sequence selected during beam search, while the green result denotes the ground-truth candidate.
Together, these examples show three typical outcomes: direct success by dual-guided generation, correction by dense vector-based reranking, and a remaining hard case involving fine-grained multimodal evidence matching.

In the first case, DrIG retrieves the ground-truth image at rank 1 during beam search.
The generated identifier follows the correct branch, and the candidate branch ending with token 173 receives an additional set-based relevance contribution.
This example demonstrates the benefit of the dual-role identifier when the sequential decoding score and the set-based global prior are aligned.
The autoregressive decoder provides a valid and semantically plausible identifier path, while the set-based role further strengthens the globally relevant branch, allowing the correct candidate to remain preferred over competing prefixes.

The second case shows an error made by beam search that is corrected by reranking.
During generation, the ground-truth text candidate remains in the top-ranked candidate list, but it is placed behind a competing branch because its combined decoding score is still lower.
Specifically, although the correct branch ending with token 139 receives an additional set-based relevance contribution, it does not surpass the competing branch ending with token 140 during the generative stage.
However, because the correct candidate is preserved within the generated top-\(k\) set, dense vector-based reranking can compare candidates in the continuous representation space and promote the ground-truth result to rank 1.
This case supports the motivation of the hybrid retrieval strategy: generative retrieval provides an efficient candidate-generation mechanism, while dense reranking compensates for residual ranking errors caused by discrete identifier decoding.

The third case highlights a remaining limitation.
DrIG correctly infers the target modality as an image--text pair and retrieves candidates that are visually related to the query image.
However, the task requires simultaneous alignment between visual evidence and fine-grained textual information.
Although the retrieved candidates are visually similar railway-station examples, the model fails to distinguish the correct textual evidence, such as the specific answer about the opening year.
The ground-truth branch associated with token 189 receives an additional set-based relevance contribution, but it still does not dominate the competing branch associated with token 187.
Moreover, dense vector-based reranking is unable to move the ground-truth candidate to the top position, suggesting that both the discrete identifier and the dense reranker struggle when the distinction depends on subtle multimodal evidence.

Overall, the case studies provide qualitative support for the proposed design.
The dual-role identifier can guide beam search toward globally relevant candidates and reduce prefix-level decoding errors, while hybrid reranking can further correct cases where the ground-truth candidate is generated but not ranked first.
At the same time, the failure case indicates that DrIG still has difficulty with retrieval scenarios requiring precise joint reasoning over visual content and textual evidence.
Improving fine-grained multimodal grounding and strengthening reranking for image--text candidates remain important directions for future work.

\section{Conclusion and Future Work}
\label{sec:conclusion}
In this paper, we proposed DrIG, a generative framework for universal multimodal retrieval with dual-role identifiers.
The key idea is to assign each candidate a single residual-quantized identifier and reuse it in two complementary ways.
In its sequential role, the identifier is generated autoregressively under a Trie constraint, enabling efficient retrieval over a discrete candidate space.
In its set-based role, the same identifier tokens provide a prefix-independent global relevance prior, which helps guide constrained beam search beyond local token-level likelihoods.
By combining these two roles during inference, DrIG reduces the risk that relevant candidates are discarded because of early prefix-level pruning. To address the information loss introduced by discrete quantization, we further introduced a hybrid retrieval strategy.
DrIG first generates a compact list of candidate identifiers and then optionally reranks the corresponding candidates using continuous embedding similarity.
This design allows the generative retriever to preserve its scalability advantage while using dense representations to recover fine-grained semantic distinctions that may be weakened during quantization.
In addition, query augmentation through query--target interpolation and a discriminative ranking objective further improve the robustness and retrieval-awareness of decoder training.

Experiments on M-BEIR demonstrate that DrIG consistently outperforms the generative universal multimodal retrieval baseline GENIUS under both local-pool and global-pool settings.
Additional text-to-image experiments on Flickr30K and MSCOCO further show that DrIG is effective beyond the heterogeneous M-BEIR benchmark.
The ablation studies confirm the importance of contrastive identifier learning, Trie-constrained decoding, query augmentation, the set-based relevance prior, the ranking objective, and modality-aware quantization.
The efficiency and sensitivity analyses further show how codebook configuration, beam size, reranking depth, decoder backbone, and prior weight affect the effectiveness--efficiency trade-off.
Overall, these results suggest that dual-role identifiers provide a promising direction for scalable generative multimodal retrieval, especially when combined with lightweight dense vector-based reranking.

Several directions remain for future work.
First, although DrIG substantially improves generative retrieval effectiveness, a gap remains between hybrid generative retrieval and the strongest dense or LMM-based retrievers on some knowledge-intensive and text-centric tasks.
Developing more expressive identifiers, better score calibration mechanisms, or stronger reranking strategies may further reduce this gap.
Second, the current framework follows a staged training pipeline in which representation learning, identifier construction, set-based scoring, and decoder training are optimized separately.
A more tightly coupled or end-to-end optimization strategy could further align discrete identifier learning with the final retrieval objective.
Third, generative retrieval systems must support dynamic candidate collections in practical applications.
Future work should investigate how DrIG can efficiently handle candidate insertion, deletion, and identifier updates without expensive retraining.
Finally, it would be valuable to evaluate DrIG in larger and more diverse multimodal retrieval scenarios, such as web-scale image--text retrieval, video retrieval, and retrieval-augmented multimodal generation.
\bibliographystyle{ACM-Reference-Format}
\bibliography{reference}
\clearpage
\appendix
\section*{Appendix}
\section{M-BEIR Instructions}
\label{app:mbeir_instructions}

This section shows the natural-language instructions used for each retrieval task type over M-BEIR. These instructions specify the retrieval intent and target modality, allowing a unified model to handle heterogeneous query--candidate 
formats, such as text, image, and image--text pairs. Table~\ref{tab:mbeir_instructions} 
summarizes the instruction templates used in our experiments.

\begin{table}[htbp]
\centering
\caption{Summary of the M-BEIR instructions.}
\label{tab:mbeir_instructions}
% 手动定义字号：[字体大小]{行间距}
\fontsize{7pt}{7.5pt}\selectfont 
\renewcommand{\arraystretch}{0.9} % 紧凑行高
\resizebox{0.8\textwidth}{!}{%
\begin{tabularx}{\columnwidth}{@{} l l X @{}} % 使用 X 自动换行
\toprule
\textbf{Task} & \textbf{Dataset} & \textbf{Instruction} \\
\midrule
\multirow{12}{*}{$q^t \rightarrow c^i$} 
& \multirow{4}{*}{VisualNews} 
& Identify the news-related image in line with the described event. \\
& & Display an image that best captures the following caption from the news. \\
& & Based on the caption, provide the most fitting image for the news story. \\
& & I want you to retrieve an image of this news caption. \\
\cmidrule{2-3}
& \multirow{4}{*}{MSCOCO} 
& Find me an everyday image that matches the given caption. \\
& & Identify the image showcasing the described everyday scene. \\
& & I want to retrieve an image of this daily life description. \\
& & Show me an image that best captures the following common scene description. \\
\cmidrule{2-3}
& \multirow{4}{*}{Fashion200K} 
& Based on the following fashion description, retrieve the best matching image. \\
& & Match the provided description to the correct fashion item photo. \\
& & Identify the fashion image that aligns with the described product. \\
& & You need to identify the image that corresponds to the fashion product description provided. \\
\midrule
\multirow{4}{*}{$q^t \rightarrow c^t$} 
& \multirow{4}{*}{WebQA} 
& Retrieve passages from Wikipedia that provide answers to the following question. \\
& & You have to find a Wikipedia paragraph that provides the answer to the question. \\
& & I want to find an answer to the question. Can you find some snippets that provide evidence from Wikipedia? \\
& & I'm looking for a Wikipedia snippet that answers this question. \\
\midrule
\multirow{8}{*}{$q^t \rightarrow (c^i, c^t)$} 
& \multirow{4}{*}{EDIS} 
& Find a news image and its caption that match the provided caption. \\
& & Identify the news photo and its caption for the given caption. \\
& & Can you pair the provided caption with the right image and its caption? \\
& & I'm looking for an image and its caption that aligns with the provided caption. \\
\cmidrule{2-3}
& \multirow{4}{*}{WebQA} 
& Find a Wikipedia image and its description that answers the given question. \\
& & Provide me with an image and its description from Wikipedia to answer the given question. \\
& & I want to know the answer to the given question. Please find the related Wikipedia image and its description for me. \\
& & You need to retrieve an evidence image with its description from Wikipedia to address the given question. \\
\midrule
\multirow{12}{*}{$q^i \rightarrow c^t$} 
& \multirow{4}{*}{VisualNews} 
& Find a caption for the news in the given photo. \\
& & Based on the shown image, retrieve an appropriate news caption. \\
& & Provide a news-related caption for the displayed image. \\
& & I want to know the caption for this news image. \\
\cmidrule{2-3}
& \multirow{4}{*}{MSCOCO} 
& Find an image caption describing the following everyday image. \\
& & Retrieve the caption for the displayed day-to-day image. \\
& & Can you find a caption talking about this daily life image? \\
& & I want to locate the caption that best describes this everyday scene image. \\
\cmidrule{2-3}
& \multirow{4}{*}{Fashion200K} 
& Find a product description for the fashion item in the image. \\
& & Based on the displayed image, retrieve the corresponding fashion description. \\
& & Can you retrieve the description for the fashion item in the image? \\
& & I want to find a matching description for the fashion item in this image. \\
\midrule
\multirow{4}{*}{$q^i \rightarrow c^i$} 
& \multirow{4}{*}{NIGHTS} 
& Find a day-to-day image that looks similar to the provided image. \\
& & Which everyday image is the most similar to the reference image? \\
& & Find a daily life image that is identical to the given one. \\
& & You need to identify the common scene image that aligns most with this reference image. \\
\midrule
\multirow{8}{*}{$(q^i, q^t) \rightarrow c^t$} 
& \multirow{4}{*}{OVEN} 
& Retrieve a Wikipedia paragraph that provides an answer to the given query about the image. \\
& & Determine the Wikipedia snippet that identifies the visual entity in the image. \\
& & I want to find a paragraph from Wikipedia that answers my question about this image. \\
& & You have to find a Wikipedia segment that identifies this image's subject. \\
\cmidrule{2-3}
& \multirow{4}{*}{InfoSeek} 
& Retrieve a Wikipedia paragraph that provides an answer to the given query about the image. \\
& & Determine the Wikipedia snippet that matches the question of this image. \\
& & I want to find a paragraph from Wikipedia that answers my question about this image. \\
& & You have to find a Wikipedia segment that answers the question about the displayed image. \\
\midrule
\multirow{8}{*}{$(q^i, q^t) \rightarrow c^i$} 
& \multirow{4}{*}{FashionIQ} 
& Find a fashion image that aligns with the reference image and style note. \\
& & With the reference image and modification instructions, find the described fashion look. \\
& & Given the reference image and design hint, identify the matching fashion image. \\
& & I'm looking for a similar fashion product image with the described style changes. \\
\cmidrule{2-3}
& \multirow{4}{*}{CIRR} 
& Retrieve a day-to-day image that aligns with the modification instructions of the provided image. \\
& & Pull up a common scene image like this one, but with the modifications I asked for. \\
& & Can you help me find a daily image that meets the modification from the given image? \\
& & I'm looking for a similar everyday image with the described changes. \\
\midrule
\multirow{8}{*}{$(q^i, q^t) \rightarrow (c^i, c^t)$} 
& \multirow{4}{*}{OVEN} 
& Retrieve a Wikipedia image-description pair that provides evidence for the question of this image. \\
& & Determine the Wikipedia image-snippet pair that clarifies the entity in this picture. \\
& & I want to find an image and subject description from Wikipedia that answers my question about this image. \\
& & I want to know the subject in the photo. Can you provide the relevant Wikipedia section and image? \\
\cmidrule{2-3}
& \multirow{4}{*}{InfoSeek} 
& Retrieve a Wikipedia image-description pair that provides evidence for the question of this image. \\
& & Determine the Wikipedia image-snippet pair that matches my question about this image. \\
& & I want to find an image and subject description from Wikipedia that answers my question about this image. \\
& & I want to address the query about this picture. Please pull up a relevant Wikipedia section and image. \\
\bottomrule
\end{tabularx}
}
\end{table}

\end{document}